\documentclass{aa}  

\usepackage{graphicx}

\usepackage{txfonts}
\usepackage{hyperref}

\usepackage{amsmath}
\usepackage{comment}
\usepackage{cleveref}
\usepackage{subfigure}
\usepackage{xcolor}

\usepackage{savesym}
\savesymbol{tablenum}
\usepackage{siunitx}
\restoresymbol{SIX}{tablenum}

\begin{document}

    \title{Three-temperature radiation hydrodynamics with PLUTO}

   \subtitle{Thermal and kinematic signatures of accreting protoplanets}

   \author{Dhruv Muley \inst{1}
        \and Julio David Melon Fuksman \inst{1}
          \and Hubert Klahr \inst{1}
          }

   \institute{Max-Planck-Institut f\"ur Astronomie, Königstuhl 17, Heidelberg, DE 69117\\
              \email{muley@mpia.de}
             }

   \date{Received 26 February 2024; accepted 3 May 2024}

 
  \abstract{
  In circumstellar disks around young stars, the gravitational influence of nascent planets produces telltale patterns in density, temperature, and kinematics. To better understand these signatures, we first performed 3D hydrodynamical simulations of a 0.012 $M_{\odot}$ disk, with a Saturn-mass planet orbiting circularly in-plane at 40 au. We tested four different disk thermodynamic prescriptions (in increasing order of complexity, local isothermality, $\beta$-cooling, two-temperature radiation hydrodynamics, and three-temperature radiation hydrodynamics), finding that $\beta$-cooling offers a reasonable approximation for the three-temperature approach when the planet is not massive or luminous enough to substantially alter the background temperature and density structure. Thereafter, using the three-temperature scheme, we relaxed this assumption, simulating a range of different planet masses (Neptune-mass, Saturn-mass, Jupiter-mass) and accretion luminosities (0, $10^{-3} L_{\odot}$) in the same disk. Our investigation revealed that signatures of disk-planet interaction strengthen with increasing planet mass, with circumplanetary flows becoming prominent in the high-planet-mass regime. Accretion luminosity, which adds pressure support around the planet, was found to weaken the midplane Doppler-flip, potentially visible in optically thin tracers like C$^{18}$O, while strengthening the spiral signature, particularly in upper disk layers sensitive to thicker lines, like those of $^{12}$CO.  
}


   \maketitle
%

\section{Introduction}

The high spectral and spatial resolution of the Atacama Large Millimeter Array (ALMA) have made it possible to accurately probe temperatures and velocities at the $\tau = 1$ surfaces of various molecular lines, such as those associated with $^{12}$CO, $^{13}$CO, and C$^{18}$O \citep{Pinte2023}. Observations in systems such as HD 163296 \citep{Pinte2018}, HD 97048 \citep{Pinte2019}, TW Hya \citep{Teague2019,Teague2022}, CQ Tau \citep{Woelfer2021}, and Elias 2-24 \citep{Pinte2023b}---with the background temperature and (sub-)Keplerian velocity profiles subtracted off---have revealed localized velocity kinks, as well as large-scale spiral structures in temperature and velocity. Numerical \citep[e.g.,][]{Perez2018} and analytical \citep[e.g.,][]{Bollati2021} studies indicate that such signatures are consistent with those caused by spiral wakes \citep{Goodman2001} launched at Lindblad resonances in the disk, where the Doppler-shifted planetary forcing frequency equals the local epicyclic frequency \citep[e.g.,][]{Goldreich1978,Goldreich1979}. 

For computational efficiency and ease of interpretation, simulations of disk-planet interaction have historically used a 2D, vertically-integrated approach, with the gas following a locally isothermal equation of state. But as the quality of observations improve, it has become increasingly necessary to account for more detailed disk structure and thermodynamics in order to offer a meaningful comparison. \cite{Zhu2012} and \cite{Lubow2014} simulated 3D adiabatic disks, discovering additional spirals excited at distinct ``buoyancy resonances'' where the Doppler-shifted planetary forcing frequency equals the Brunt-Väisälä frequency. \cite{LoboGomes2015} simulated disk-planet interaction with cooling in 2D, running their simulations to gap-opening timescales with an emphasis on vortex formation at the outer pressure bump formed by the planetary gap. \cite{Zhu2015} performed global 3D simulations of planet-driven spirals with so-called ``$\beta$-cooling'' to a background temperature structure (with $\beta$ being the ratio of cooling time to local dynamical time, $\Omega_K^{-1}$). \cite{Juhasz2018} performed 3D locally isothermal prescriptions, but with a vertically stratified temperature. \cite{Miranda2020,Miranda2020b} performed high-resolution, 2D simulations of spirals with $\beta$-cooling, studying the details of angular momentum transport as cooling times varied from short (isothermal) to long (adiabatic), relative to the local dynamical time. The 3D simulations of \cite{Muley2021} incorporated $\beta$-cooling along with a vertically stratified temperature structure obtained from radiative-transfer simulations, while those of \cite{Bae21} went a step further and computed $\beta$-cooling timescales at each point in the disk, based on radiative diffusion and gas-grain collision times. These studies conclude that Lindblad spirals propagate through the disk for $\beta \ll 1$ (isothermal limit) or $\beta \ll 1$ (adiabatic limit), but damp close to the planet location for $\beta \approx 1$. Temperature stratification also causes the pitch angle and morphology of these spirals to deviate from the 2D, vertically-averaged expectation, particularly at the high altitudes amenable to observations.

In addition to parametrized cooling, radiation-hydrodynamic techniques have also been used in the context of disk-planet interaction. A number of works \citep[e.g.,][]{Kley2009,Lega2014,Fung2017,Chrenko2020,Yun2022} have concentrated on planetary torques and migration, while others \citep[e.g.,][]{Klahr2006,Szulagyi2017,Chrenko2019} have focused on flows in the circumplanetary region. These have typically employed a one-temperature \citep[1T;][]{Kley1989} scheme---in which gas and radiation are assumed to have the same temperature---or a more involved two-temperature \citep[2T;][]{Bitsch2013a} scheme in which matter and radiation have separate temperatures, coupled by opacity. For the transport of radiation, these simulations have used flux-limited diffusion \cite[FLD;][]{Levermore1981} approaches to solving the governing equations, including various parametrizations for stellar irradiation. The high-resolution 2D simulations of \cite{Ziampras2023}, using the two-temperature FLD scheme outlined in \citep{Ziampras2020}, gave more emphasis to the morphologies of spirals themselves. Their work found that transport of energy across the spiral shock \citep{Ensman1994,Commercon2011} shifts and broadens the spiral temperature perturbation in ways that an inherently local, $\beta$-cooling approximation cannot. Concurrently, \cite{Muley2023} ran 3D simulations with M1 radiation transport \citep{Levermore1984,MelonFuksman2019} and ray-traced stellar irradiation, as a test of their ``three-temperature'' (3T) approach, in which gas, grains, and radiation coupled by collisions and opacity. In both two- and three-temperature simulations, they found weaker pre-shock heating than in \cite{Ziampras2023}, potentially attributable to a lower disk mass enabling efficient vertical cooling (Ziampras 2023, private communication).

In this work, we build on these previous disk-planet simulations using our 3T method, with the aim of better connecting kinematic and thermal spiral signatures to the properties of the planets driving them. In Section 2, we describe our methods, including our thermodynamic prescriptions, treatment of planetary accretion luminosity, and disk initial conditions. In Section 3, we discuss the spiral structure, flow patterns, and background temperature created by a non-accreting, Saturn-mass planet orbiting at 40 au in the disk, testing disk-planet interaction under four different thermodynamic prescriptions (local isothermality, physically motivated $\beta$-cooling, two-temperature radiation hydrodynamics, and three-temperature radiation hydrodynamics). In Section 4, we run three-temperature simulations only, measuring the effects of changing mass and accretion luminosity. In Section 5, we present polar cuts of temperature and sky-projected velocity high in our simulated disk, and comment on the observational implications. Finally, in Section 6, we summarize and conclude our work.

\section{Methods}
\subsection{Basic equations}\label{sec:basic_equations}
For our study of spiral arms, we use a version of the \texttt{PLUTO} hydrodynamical code \citep{Mignone2007}, modified to solve the equations of radiation hydrodynamics \citep{MelonFuksman2019,MelonFuksman2021} with an additional dust internal energy field \citep{Muley2023}. This field interacts thermally with the gas and radiation field, but passively traces the same velocity field as the gas without any back-reaction (implying a Stokes number $\mathrm{St} \ll 1$, as well as a globally constant dust-to-gas ratio $f_d \ll 1$):

\begin{subequations}\label{eq:radhydro}
\begin{equation}
    \frac{\partial \rho}{\partial t} + \nabla \cdot (\rho \vec{v}) = 0
\end{equation}
\begin{equation}
    \frac{\partial (\rho \vec{v})}{\partial t} + \nabla \cdot (\rho \vec{v} \vec{v}) = -\nabla p - \rho \nabla \Phi + \vec{S}_m + \vec{G}
\end{equation}
\begin{equation}
    \frac{\partial E_g}{\partial t} + \nabla \cdot (E_g \vec{v}) = -\nabla \cdot ((p + \rho \Phi) \vec{v}) + S_m + X_{gd} + cG_g  + S^{\rm irr}_g
\end{equation}
\begin{equation}
\label{eq:ed_evo}
    \frac{\partial E_d}{\partial t} + \nabla \cdot (E_d \vec{v}) = -X_{gd} + cG_d + S^{\rm irr}_d
\end{equation}
\begin{equation}
\label{eq:er_evo}
    \frac{\partial E_r}{\partial t} + \hat{c}\nabla \cdot \vec{F}_r = -\hat{c}(G_g + G_d)
\end{equation}
\begin{equation}
\label{eq:flux_evo}
    \frac{\partial \vec{F}_r}{\partial t} + \hat{c}\nabla \cdot \mathbf{P}_r = -\hat{c}\vec{G}
\end{equation}
\end{subequations}
in which $\rho, \vec{v}, p$ represent the gas density, velocity, and pressure respectively. $\rho_d$ is the dust density, while $f_d$ is the dust-to-gas ratio, $\Phi$ is the gravitational potential, and $\{E_g, E_d, E_r\}$ are total energy densities for gas, dust, and radiation respectively. $\vec{S}_m$ represents parabolic source terms such as $\alpha$-viscosity,  $X_{\rm gd}$ represents energy exchange between gas and dust. $\vec{F}_r$ is the radiative flux, while $\vec{P}_r$ is the radiation pressure. $G_g$, $G_d$, and $\vec{G}$ terms represent opacity-mediated interaction between the gas, dust and radiation respectively. $S_{\rm g}^{\rm irr}$ and $S_{\rm d}^{\rm irr}$ represent, respectively, gas and dust absorption of stellar irradiation. $c$ represents the speed of light, whereas the $\hat{c}$ term is a ``reduced speed of light'' \citep{Gnedin2001} which enables longer timesteps, but must nevertheless exceed all hydrodynamic velocities relevant to the problem \citep{Skinner2013}. 

The opacity source terms are given by:
\begin{subequations}
\begin{equation}
\begin{split}
    G_g \equiv &-\rho \kappa_{g}(a_r T_g^4 - E_r) \\
&- \rho (2 \kappa_{g} - \chi_g) \vec{\beta} \cdot \vec{F}_r -\rho \chi_{g} \vec{\beta} \cdot (E_r \vec{\beta} + \vec{\beta} \cdot \mathbf{P}_r)
\end{split}
\end{equation}
\begin{equation}
\begin{split}
    G_d \equiv &-\rho \kappa_{d} f_d (a_r T_d^4 - E_r) \\ &- \rho f_d (2 \kappa_{d} - \chi_d) \vec{\beta} \cdot \vec{F}_r -\rho f_d \chi_{d} \vec{\beta} \cdot (E_r \vec{\beta} + \vec{\beta} \cdot \mathbf{P}_r)
\end{split}
\end{equation}
\begin{equation}
\begin{split}
    \vec{G}_g \equiv & \rho \chi_g \vec{F}_r -\rho \kappa_{g}(a_r T_g^4 - E_r) \vec{\beta}\\
&- 2 \rho \kappa_{g} (\vec{\beta} \cdot \vec{F}_r) \vec{\beta} -\rho \chi_{g} (E_r \vec{\beta} + \vec{\beta} \cdot \mathbf{P}_r)
\end{split}
\end{equation}
\begin{equation}
\begin{split}
    \vec{G}_d \equiv & \rho f_d \chi_d \vec{F}_r -\rho f_d \kappa_{d}(a_r T_d^4 - E_r) \vec{\beta}\\
&- 2 \rho f_d \kappa_{d} (\vec{\beta} \cdot \vec{F}_r) \vec{\beta} -\rho f_d \chi_{d} (E_r \vec{\beta} + \vec{\beta} \cdot \mathbf{P}_r)
\end{split}
\end{equation}
\end{subequations}
where $\vec{G} \equiv \vec{G}_g + \vec{G}_d$ and $\vec{\beta} \equiv \vec{v}/c$ (not to be confused with the $\beta$-cooling timescale).

The solution strategy for these equations is described thoroughly in the aforementioned articles \citep{MelonFuksman2019,MelonFuksman2021,Muley2023}, and we recapitulate the most relevant details here. Equations (1) are divided into radiation and hydrodynamic subsystems, which are solved using a Strang split (half-step radiation, full-step hydro, half-step radiation). For the radiation subsystem, the Courant-Friedrichs-Levy (CFL) condition imposed by $\hat{c}$ is far more stringent than that imposed on the hydrodynamic subsystem by the sound speed $c_s$, so the radiation half-step is in turn divided into a number of substeps. Each individual substep is handled using an implicit-explicit (IMEX) strategy \citep{Pareschi2005}: radiation transport terms are computed explicitly using the M1 formalism to handle both the optically thick diffusion and optically thin beaming limits \citep{Levermore1984}, whereas the stiff $G_g$, $G_d$, $\vec{G}$, and $X_{\rm gd}$ source terms are solved implicitly with a Newton-Raphson method. The stellar irradiation source term $S_{\rm d}^{\rm irr}$ is computed explicitly by ray-tracing at the start of each hydrodynamic timestep (with $S_{\rm g}^{\rm irr}$ set to zero in the current study); because it depends only on the density distribution, it is sufficient to simply incorporate it into the initial guess of our Newton scheme, without updating it at each iteration.

We assume that the gas follows an ideal equation of state,
\begin{equation}\label{eq:prs_ideal}
    p = \rho_g k_B T_g/\mu m_H
\end{equation}
where $k_B$ is the Boltzmann constant, $T_g$ the gas temperature, $\mu$ the mean molecular weight, and $m_H$ the mass of a hydrogen atom. The adiabatic index $\gamma \equiv \partial \ln p/\partial \ln \rho$ is a constant, implying an \textit{internal} energy density
\begin{equation}\label{eq:eint_ideal}
    \xi_g = p/(\gamma - 1)\,.
\end{equation}
We note that realistic equations of state---in which $\gamma$ varies with temperature as rotational, vibrational, and dissociation modes of para- and ortho-hydrogen are activated \citep{Decampli1978,Boley2007,Boley2013}---would change Equations \ref{eq:prs_ideal} and \ref{eq:eint_ideal}, as well as the gas-dust energy exchange term in Equation \ref{eq:x_gd}, and the gas-dust coupling timescale in Equation \ref{eq:tcool}, in the following section.

\subsubsection{Three-temperature simulations}
For our three-temperature simulations, we solve the full set of Equations \ref{eq:radhydro}, and define the dust-gas collision term
\begin{equation}\label{eq:x_gd}
    X_{gd} \equiv t_c^{-1}(r_{gd} \xi_d - \xi_g)
\end{equation}
in which $t_c$ is the dust-gas thermal coupling time, $\xi_d = E_d$ represents the dust internal energy (equivalent in our scheme, but in general different if dust dynamics were to be accounted for), and $r_{gd} = c_g/ f_d c_d$ is the ratio of heat capacity per unit volume between gas and dust. $c_d$ is the specific heat capacity of dust, while $c_g \equiv k_B/\mu m_H (\gamma - 1)$ is that of the gas. We compute the gas cooling time $t_c$ as a function of the dust-gas stopping time $t_s$, calculated in the Epstein regime \citep{BurkeHollenbach83,Speedie2022}

\begin{equation}
\label{eq:tcool}
    t_c = \frac{2/3}{\gamma - 1} f_d^{-1} t_s \eta^{-1}
\end{equation}
where we set the ``accommodation coefficient'' $\eta$ to unity. 

\subsubsection{Two-temperature simulations}
In our studies of the ``two-temperature'' regime of traditional radiation hydrodynamics, in which dust and gas are perfectly well-coupled, we set the stopping time to an artificially low $t_s = 10^{-10} {\rm \ yr}$.

\subsubsection{$\beta$-cooling simulations}\label{sec:beta_cooling}

For beta-cooling simulations, we replace $X_{\rm gd}$ with a term of the form
\begin{equation}
    t_{\rm rel}^{-1} \frac{\rho k_B}{\mu m_p (\gamma - 1)} \left(T_g - T_{\rm g, 0}(r, \theta)\right)
\end{equation}

where $T_{g, 0}$ is the initial condition for gas temperature (see Section \ref{sec:setup} for more details), $\vec{x}$ is a position in the disk and $t_{\rm rel}$ can in general be a function of any primitive variables. We ignore the $G_g$ and $S_g^{\rm irr}$ terms, while eliminating Equations \cref{eq:ed_evo,eq:er_evo,eq:flux_evo} entirely. Following \cite{Bae21} and \cite{MelonFuksman2023}, we set the thermal relaxation/cooling time
\begin{equation}\label{eq:t_rel}
    t_{\rm rel} = t_c + t_{\rm rad}
\end{equation}
where $t_c$ is defined as in equation \ref{eq:tcool}, and $t_{\rm rad}$ is a radiative cooling timescale incorporating both the optically thick diffusion and optically thin free-streaming limits:
\begin{equation}
    t_{\rm rad} \equiv \max (\lambda_{\rm thin}^2, \lambda_{\rm diff}^2)/D
\end{equation}
where the radiative diffusion coefficient $D \equiv 4 c a_r T_g^3 / 3 c_g \kappa_R \rho^2$. The effective optically thin cooling length scale $\lambda_{\rm thin} = (3 \kappa_R \kappa_P \rho^2)^{-1/2}$, while the thick diffusion length is assumed to equal the local scale height, $\lambda_{\rm diff} \equiv H = c_{s, \rm iso} \Omega^{-1}$, where $\Omega$ is the local Keplerian orbital frequency and $c_{s, \rm, iso} \equiv \sqrt{p/\rho}$ the ``isothermal sound speed''.
\subsubsection{Locally isothermal simulations}
To test the locally isothermal case, we simply run a $\beta$-cooling simulation with $t_{\rm rel} = 10^{-10} {\rm \ yr}$ everywhere in the disk. 

\subsection{Implementation of accretion luminosity}\label{sec:acclum}
We implement accretion luminosity as part of the irradiation source term, $S_{\rm d}^{\rm irr}$. This allows us to use the IMEX strategy discussed in Section \ref{sec:basic_equations} to deposit, exchange, and transport large amounts of energy into each grid cell, avoiding the numerical instabilities that would arise from an explicit approach. In this work, we model only the luminosity, without removing any mass from the grid or adding any to the planet.

We interpolate the luminosity onto the grid using a triangular-shaped cloud \citep[TSC; e.g.,][]{Eastwood1986} approach, in which the kernel takes the form
\begin{equation}
    \begin{split}
        \psi(\vec{x}, \vec{x}_p) &= \max\left[\left(1 - \left|\frac{r - r_p}{\Delta r_{i_p}}\right|\right)\left(1 - \left|\frac{\theta - \theta_p}{\Delta \theta_{j_p}}\right|\right)\left(1 - \left|\frac{\phi - \phi_p}{\Delta \phi_{k_p}}\right|\right), 0\right] \\
        &\times \ (r^2 \sin \theta)^{-1}
    \end{split}
\end{equation}

where $\{i_p, j_p, k_p\}$ indicate the indices in the $\{r, \theta, \phi\}$ directions of the cell in which the planet is located. The kernel intersects the cells $\{i_p \pm 1, j_p \pm 1, k_p\pm 1\}$, for a total of 27 cells; \footnote{This only strictly holds when the cell spacings in all directions are independent of the corresponding coordinate; in the $r$-direction, where $\Delta r_i \propto r_i$, it is possible for the edge of the kernel to intersect the second cell inward, $r_{i_p - 2}$. The extent of this edge is, at most, $\Delta r_{i_p} - \Delta r_{i_p - 1}$, which for our grid is always less than 1\% of the kernel half-width; the integral of $\psi$ over this intersecting edge is $< \mathcal{O}(10^{-4})$. We add this small excess to the cells $\{i_p, j_p \pm 1, k_p \pm 1\}$.} the fraction of accretion energy deposited in each cell during a timestep is equal to the integral of $\psi$ over the cell volume. TSC avoids creating spurious discontinuities in the deposited radiation field or its gradient, unlike the cloud-in-cell (CIC) or nearest grid point (NGP) methods, where moving across a cell boundary or even within a cell can cause sharp changes in these quantities.

\begin{figure}
    \centering
    \includegraphics[width=0.45\textwidth]{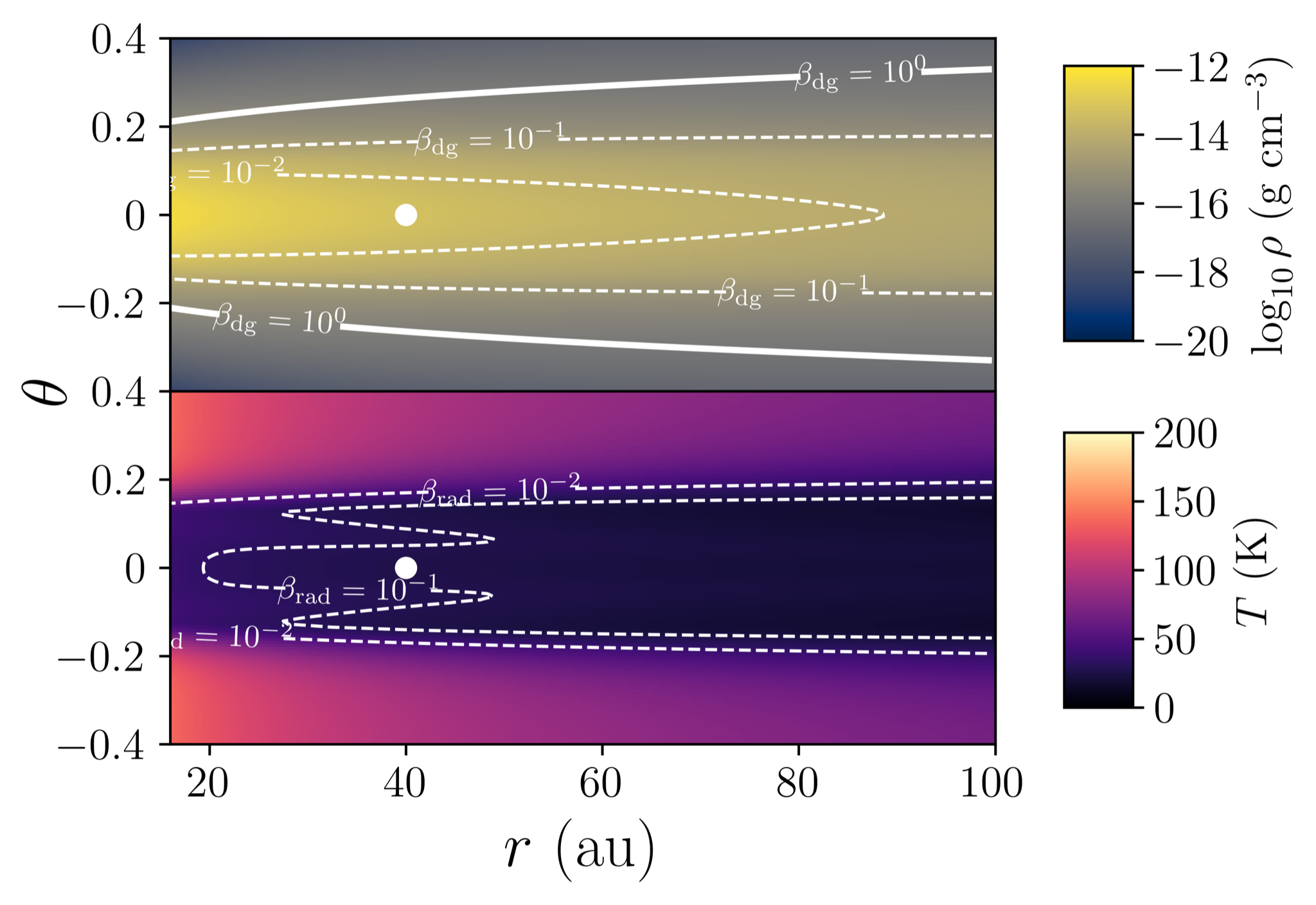}
    \caption{A plot of initial conditions for density \textit{(above)} and temperature \textit{(below)}, obtained using the iterative procedure described in Section \ref{sec:setup}. White lines indicate the corresponding cooling-time contours, computed using Equation \ref{eq:t_rel}. Due to our constant dust-to-gas ratio and assumption of small dust grains throughout the disk, we obtain shorter radiative-diffusion and gas-grain coupling times than in \cite{Bae21}.}
    \label{fig:initial_conditions}
\end{figure}

\begin{figure*}
    \centering
    \includegraphics[width=0.45\textwidth]{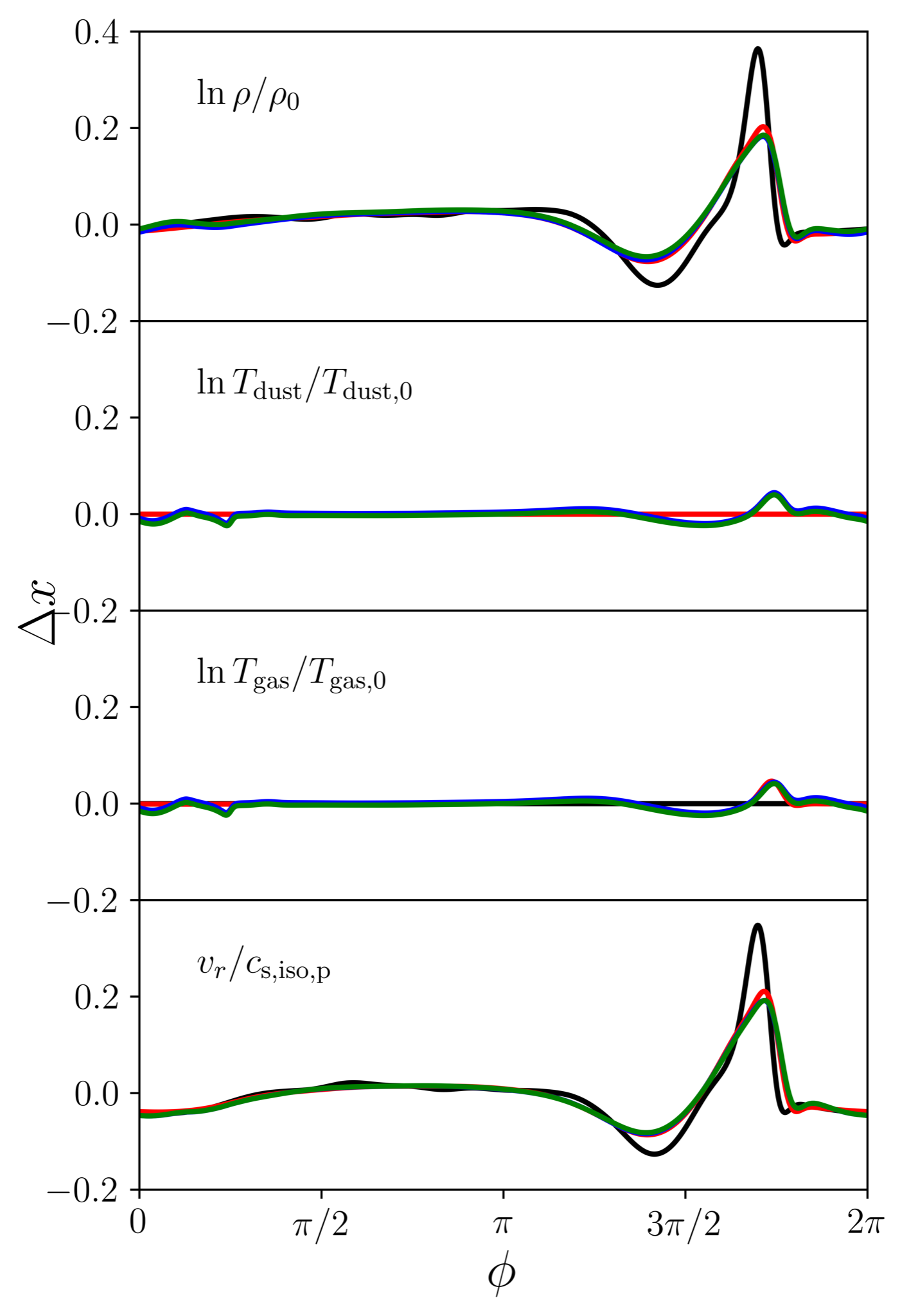}
    \includegraphics[width=0.45\textwidth]{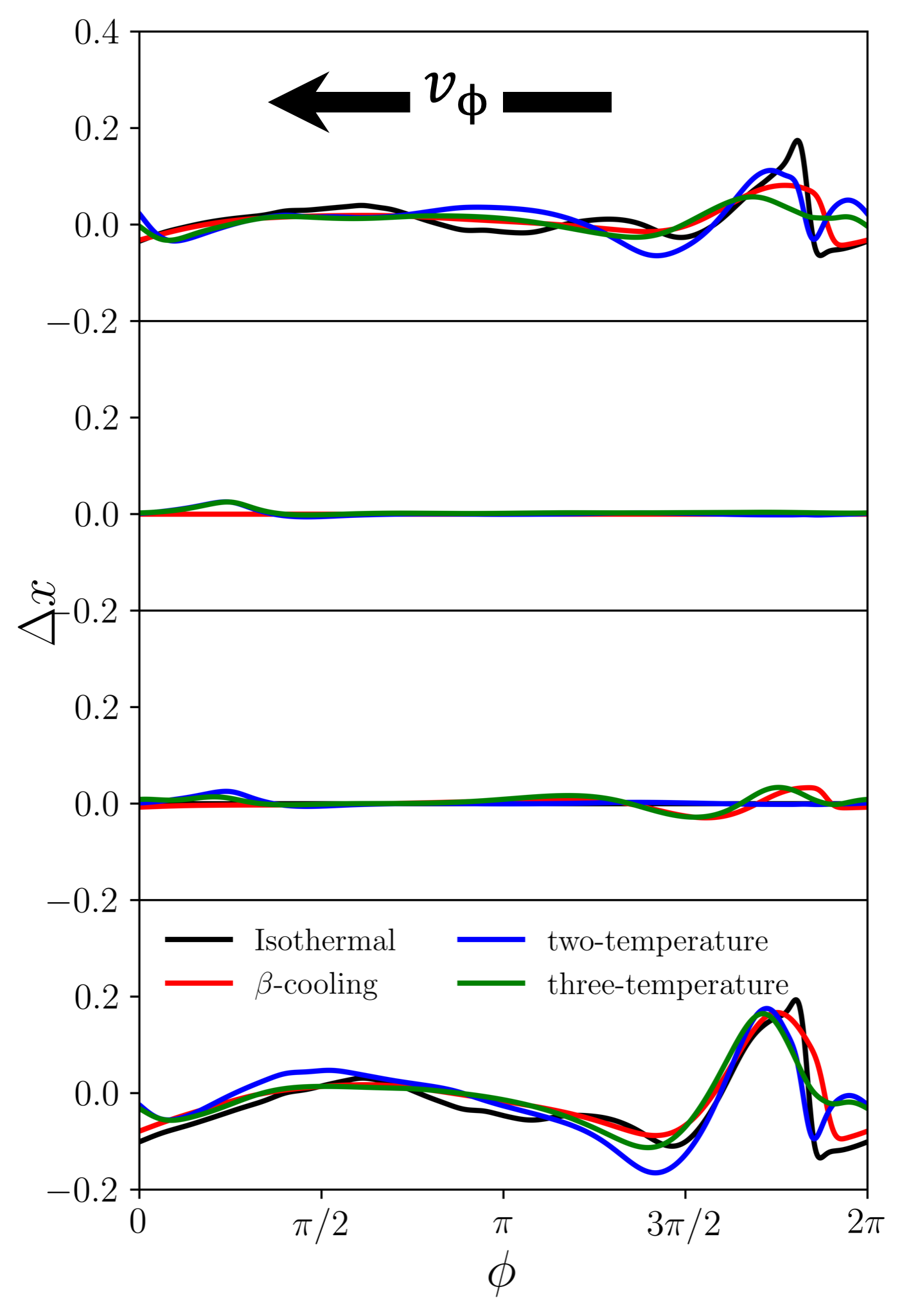}
    \caption{Relative differences in disk density, dust temperature, gas temperature, and radial velocity at $t = 2500 {\rm yr} \approx 10$ orbits, with respect to initial conditions for different physical prescriptions. We take azimuthal cuts over the whole $2\pi$ at fixed $r = 1.5 r_p$ and $\theta = 0$ \textit{(left)} and 0.2 \textit{(right)} above the midplane. Dust temperatures largely agree between the two- and three-temperature cases, whereas gas temperatures largely agree between the three-temperature and $\beta$-cooling cases. }
    \label{fig:azi_cuts_physics}
\end{figure*}

\subsection{Setup and initial conditions}\label{sec:setup}
In all simulations presented here, we assume a disk gas surface density profile of
\begin{equation}
    \Sigma_g = 200 \si{\gram\per\cm\squared} \left(\frac{R}{1 {\rm au}}\right)^{-1} \,
\end{equation}

corresponding to an $M_d = 0.012 M_{\odot}$ within our domain. As in our previous works \citep[e.g.,][]{MelonFuksman2022,MelonFuksman2024a,MelonFuksman2024b}, we approximate the behavior of the gas with an ideal equation of state, with  adiabatic index $\gamma = 1.41$ and mean molecular weight $\mu = 2.3$ \footnote{We note that within our temperature range of interest ($\approx 20-200$ K), $\gamma$ can vary between 1.3-1.66 as rotational modes of the $H_2$ molecule are activted, with the exact index depending on the assumed ortho- and para-hydrogen ratio \citep[equilibrium or some fixed fraction; see e.g.,][]{Bitsch2013}.} The total dust-to-gas ratio is assumed to be 1\%, of which we take 10\% by mass (corresponding to $f_d = 10^{-3}$) to be the small grains that we model. These consist of 62.5\% silicate and 37.5\% graphite, with a material density $\rho_{\rm gr} = 2.5 \ \si{\gram\per\cubic\cm}$ and a specific heat capacity $c_d = 0.7 \ \si{\joule\per\gram\per\kelvin}$; their sizes follow the MRN \citep{MRN77} distribution, $n(a) \propto a^{-3.5}$, with $a_{\rm min} = 5 \ \si{\nano\meter}$ and $a_{\rm max} = 250 \ \si{\nano\meter}$. Using the frequency-dependent opacities given by \cite{Krieger2020,Krieger2022}, we create tables of Planck and Rosseland opacity for this grain distribution as a function of temperature.

In order to obtain initial conditions, we employ the same iterative technique used in \cite{MelonFuksman2022}, cycling hydrostatic-equilibrium and radiative transfer calculations on a timescale ($t_{\rm iter} = 0.1{~\rm y}$) much shorter than thermal diffusion times through the disk. We run 1000 iterations in order to obtain a well-converged background profile, and use this profile as the initial condition for all simulations, regardless of the physics prescription implemented. 

For our planet, located at $r_p = 40$ au, $\theta_p = \pi/2$, and $\phi_p = \pi/4$, this profile yields a local temperature $T_p = T_{\rm g, 0}(r = 40 {\rm au}, \theta = \pi/2) = 26$ K and a scale-height ratio $(H/r)_p = 0.065$. We plot our initial conditions in Figure \ref{fig:initial_conditions}, along with contours for the gas-grain coupling time (on the upper density plot) and radiative cooling time (on the lower temperature plot), normalized by the local dynamical time to obtain effective $\beta$-values; in our $\beta$-cooling simulations we simply add $\beta_{\rm dg} + \beta_{\rm rad}$. Compared with the detailed calculations of \cite{Bae21}, who self-consistently computed grain settling, opacities, and radiative cooling rates, our assumption of globally constant dust-to-gas ratio and grain size distribution yields shorter cooling times throughout the disk. In the upper atmosphere, this is because our unsettled small grain distribution makes collisional coupling more efficient; in the midplane, the low opacities of these small grains at tens of Kelvin shortens the thermal diffusion timescale.

For all simulations presented here, we use WENO3 reconstruction, along with a van Leer limiter and shock flattening to ensure stability. Our grid resolution is $268 (r) \times 116 (\theta) \times 919 (\phi)$, logarithmically spaced in the radial and evenly spaced in the polar and azimuthal. Our domain extends between $r = \{0.4, 2.5\} \times r_p$, $\theta = \{-0.4, 0.4\} + \pi/2$, and $\phi = \{0, 2\pi\}$. Given these dimensions, our resolution yields $\sim{10}$ cells per scale height at the planet location, the same amount used in historical 2D and 3D simulations of planetary gap-opening \citep{FSC14,Fung2016}, and higher than in the 3D spiral simulations of \cite{Muley2021}, but somewhat lower than in the 2D simulations of \cite{Ziampras2023}, which contains a more detailed analysis of angular-momentum fluxes and multi-gap opening induced by spirals. We use a reduced speed of light $\hat{c} = 10^{-4} c$, which \cite{Muley2023} found to work well for analogous simulations of spirals driven by planets tens of au from their host stars.

In the two-temperature case, our test simulations show that disk columns can oscillate around the midplane, with low-order azimuthal modes undergoing linear growth. Because these modes do not manifest in the isothermal and $\beta$-cooling simulations, and appear even in the absence of a planet, we tentatively attribute them to some form of irradiation \citep{Fung2014} or self-shadowing \citep[e.g.,][and references therein]{MelonFuksman2022} instability. In the three-temperature case, this effect is naturally suppressed---we surmise due to the decoupling between gas and dust temperatures at the high altitudes where stellar irradiation is intercepted. As such, for the two-temperature simulation only, we impose rapid wave-damping zones in the upper and lower regions of the domain where $|\theta - \pi/2| > 0.3$, and defer further investigation of this phenomenon to future work. 

\section{Disk physics}

\begin{figure*}\centering
    \includegraphics[width=0.45\textwidth]{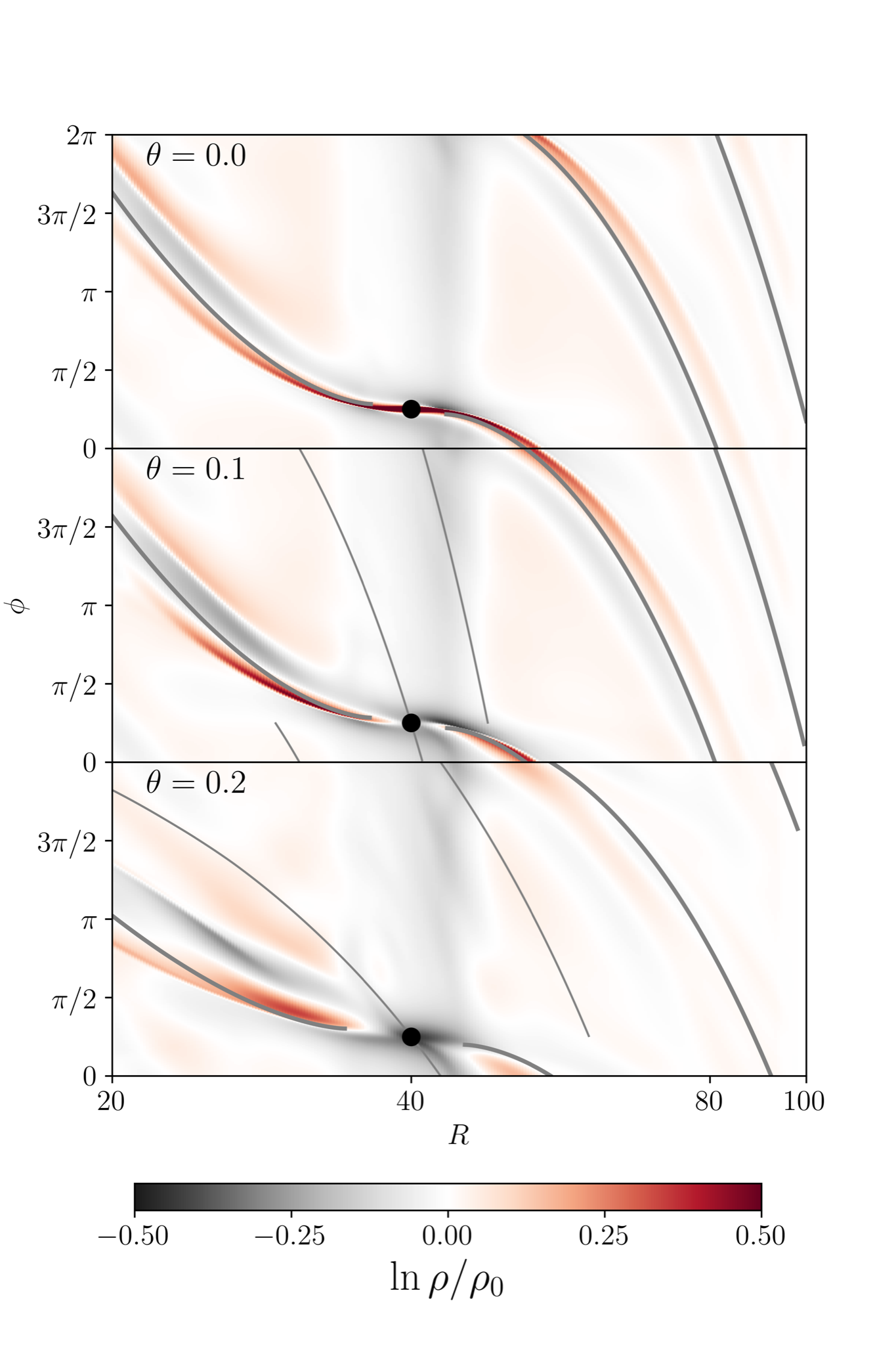}
    \includegraphics[width=0.45\textwidth]{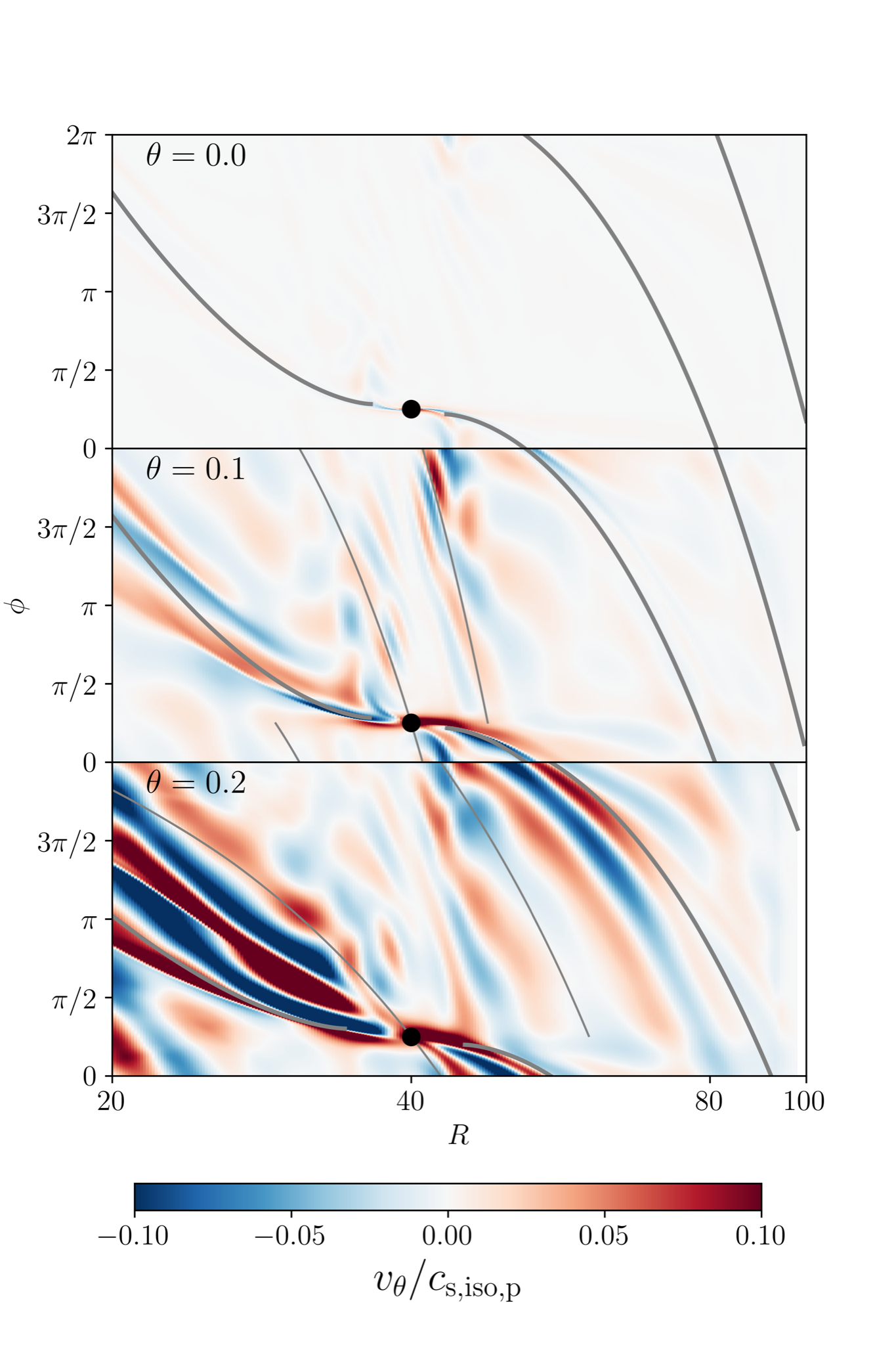}
    \caption{Plots of the density \textit{(left)}, indicating the planetary spirals, and meridional velocity \textit{(right)}, for various altitudes in the disk; phase predictions for the primary Lindblad arm (Equation \ref{eq:lindblad_phase}, thick line) and $m = 3$ buoyancy arm (Equation \ref{eq:buoyancy_phase}, thin line) are overplotted. In all panels, cylindrical radius $R \equiv r \sin \theta$ is plotted on the $x$-axis and azimuthal angle $\theta$ on the $y$-axis. Due to the relatively rapid thermal relaxation in this simulation, buoyancy spirals are markedly weaker than Lindblad spirals in all cases.}
    \label{fig:pitch_angles}
\end{figure*}
\subsection{Thermodynamic prescriptions}\label{sec:thermodynamic_test}
In Figure \ref{fig:azi_cuts_physics}, we plot perturbations in density $\rho$, gas temperature $T_g$, dust temperature $T_d$, and perturbations in radial (in center-of-mass coordinates) velocity $v_r$, for each of our four thermodynamic prescriptions (to recapitulate, local isothermality, $\beta$-cooling, two-temperature radiation hydrodynamics, and three-temperature radiation hydrodynamics). We take cuts of these variables at $r = 1.5 r_p$ at $\theta = 0$ and $\theta = 0.2$ above the midplane, at $t = 2500$ yr. \footnote{Here, and throughout our analysis, $\theta$ is implied to be with respect to the midplane angle $\pi/2$.}

In the midplane, we find that perturbations in $\rho$ and $v_r$ are closely aligned with one another in $\phi$. In agreement with previous results \citep[e.g.,][]{Zhu2015,Muley2021}, we find these perturbations strongest in the locally isothermal simulations. Among our non-isothermal runs, perturbations in $\rho$, $v_r$, and $T_g$ are similar regardless of whether a $\beta$-cooling, 2T, or 3T prescription is used. $T_g$ perturbations capture the work done by compression and expansion ($p dV$) on a fluid parcel in the disk over one thermal relaxation time. In the midplane of our fiducial disk, this is typically comparable to or shorter than the spiral-crossing time ($\beta \lesssim 1$), so the measured $T_g$ is relatively weak and somewhat offset in $\phi$ with respect to the $\rho$ perturbation (which traces the integrated compression and expansion over a fluid parcel's orbit).

Owing to the very short dust-gas coupling time $t_c$ in this region, $T_d$ closely follows $T_g$, and likewise agrees between the two- and three-temperature setups. The $\beta$-cooling simulation, which assumes a constant background temperature, is inherently unable to capture changes in $T_d$. Unlike in the 2D flux-limited diffusion simulations of \cite{Ziampras2023}, we do not observe substantial radiative heating of the pre-shock region by emission from hot, post-shock gas \citep[e.g.,][]{Ensman1994}. This may be caused by efficient cooling through the disk surface (Ziampras 2023, private communication) in our relatively low-mass, optically thin disk. 

This picture becomes somewhat more complicated at $\theta = 0.2$, where the cooling time lengthens, and becomes dominated by dust-gas decoupling rather than radiative diffusion. The $\rho$ spiral is sharp in the locally isothermal case, but weaker in the other simulations. $T_d$ agrees between the two- and three-temperature simulations, but the gas temperature differs, indicating that observational signatures observable in the upper disk gas (in, e.g., $^{12}$CO) may not be reflected in the dust.

\subsection{Lindblad and buoyancy resonances}
The observed disk signatures are, in large part, driven by resonant interactions between the disk and the planet's gravitational potential. One such interaction, studied by decades of analytical and numerical work, occurs at the \textit{Lindblad resonances}, where the Doppler-shifted planetary forcing frequency times an azimuthal integer wavenumber $m$ equals the epicyclic frequency $\kappa$ \citep[e.g.,][]{Goldreich1978,Goldreich1979,Goldreich1980,Kley2012,Bae2018a,Bae2018b}. Assuming that the disk is thin, and that the spirals are tightly wound ($k_r \gg mr^{-1}$) so the Wentzel-Kramers-Brillouin (WKB) approximation holds, we obtain the following dispersion relation:
\begin{equation}
    m^2(\Omega - \Omega_p)^2 = \kappa^2 + c_s^2 k_r^2
\end{equation}
where $\Omega_p$ is the planet's orbital frequency, and $k_r$ is the radial wavenumber of the excitation. Given that in a Keplerian disk $\kappa = \Omega \propto r^{-3/2}$, we can rewrite the above in terms of $k_r$ and $m$, and integrate to find the phase of the $m$th spiral mode:
\begin{equation}\label{eq:lindblad_phase}
    \phi_{L,m}(r) = \int_{r_m}^r \left[\frac{H(r')}{r'}\right]^{-1} \left[\left(1 - \frac{\Omega_p}{\Omega(r')}\right)^2 - m^{-2}\right]^{1/2} dr'
\end{equation}
where $r_m = (1 \mp m^{-1})^{2/3} r_p$ is the resonance location. A more detailed analysis, going beyond the WKB approximation \citep{Artymowicz1993,Papaloizou2007} shows that the peak mode strength occurs at approximately $m \gtrsim 2/(H/r)$, above which wave excitation becomes inefficient.

One can also perform a similar analysis on the the \textit{buoyancy resonances}, where the $m$-multiplied, Doppler-shifted forcing frequency equals the (vertical) Brunt-Väisälä frequency $N_z \equiv g\gamma^{-1}\left[\ln{P/\rho^{\gamma}}\right]^{1/2} $ \citep{Lubow2014}. These are an inherently 3D phenomenon, depending sensitively on the disk's thermal physics prescription. Applying the thin-disk approximation that vertical gravity $g \approx \Omega^2 z$ and assuming a background in hydrostatic equilibrium, one can simplify the above to:
\begin{equation}
    m^2(\Omega - \Omega_p)^2 = N_z^2 = \Omega^2 z \left[\frac{1}{T_{g,0}} \frac{dT_{g,0}}{dz} + \frac{\gamma - 1}{\gamma} \frac{\Omega^2 z}{c_{s,\rm iso,0}^2}\right]
\end{equation}
where $c_{\rm s, iso, 0} \equiv \sqrt{p_0/\rho_0}$ is the isothermal sound speed in the initial condition. Following \citep{Zhu2015,Bae21}, this can be used to obtain a phase angle for the spiral:
\begin{equation}\label{eq:buoyancy_phase}
    \phi_{B, m} = \pm 2\pi m (1 - \Omega_p/\Omega) \left[\frac{z}{T_{g,0}} \frac{dT_{g,0}}{dz} + \frac{\gamma - 1}{\gamma} \frac{\Omega^2 z^2}{c_{s,\rm iso,0}^2}\right]^{-1/2}
\end{equation}

In Figure \ref{fig:pitch_angles} we plot the phase angles for our three-temperature simulations, density for Lindblad spirals and meridional velocities for buoyancy spirals. Lindblad spirals are clearly visible at all altitudes, but are somewhat less tightly bound than the linear WKB theory of Equation \ref{eq:lindblad_phase} would predict, given the fiducial planet's ``thermal mass'' $q_{\rm th} \equiv (M_p/M_\odot) h_p^{-3}$ of 1.12. The deviation becomes stronger, and the spiral arm becomes more open, with increasing altitude \citep[an effect also seen in the locally isothermal, temperature-stratified simulations of][]{Juhasz2018}. The vertical velocity component of the Lindblad spirals also becomes more pronounced in the disk upper atmosphere, especially in the inner disk where scale height is lower than at the wave-launching location near the planet.

As expected, buoyancy spirals are only visible in the upper layers of the disk, but likewise deviate somewhat from the linear phase prediction from Equation \ref{eq:buoyancy_phase}. In comparison to previous works, the fact that the thermal relaxation time $\mathbf{t_{\rm rel} \lesssim N_z^{-1}}$, even at high altitudes, acts to damp the buoyant oscillations. We surmise that they could be strengthened either through substantial dust growth/settling/depletion which would increase the gas-grain coupling time $t_c$, or alternatively (especially at $\theta \approx 0.1$ from the midplane, and in the inner disk) by a high dust-to-gas ratio or disk mass, which would increase the thermal diffusion time $t_{\rm diff}$.

\begin{figure}
    \centering
    \includegraphics[width=0.45\textwidth]{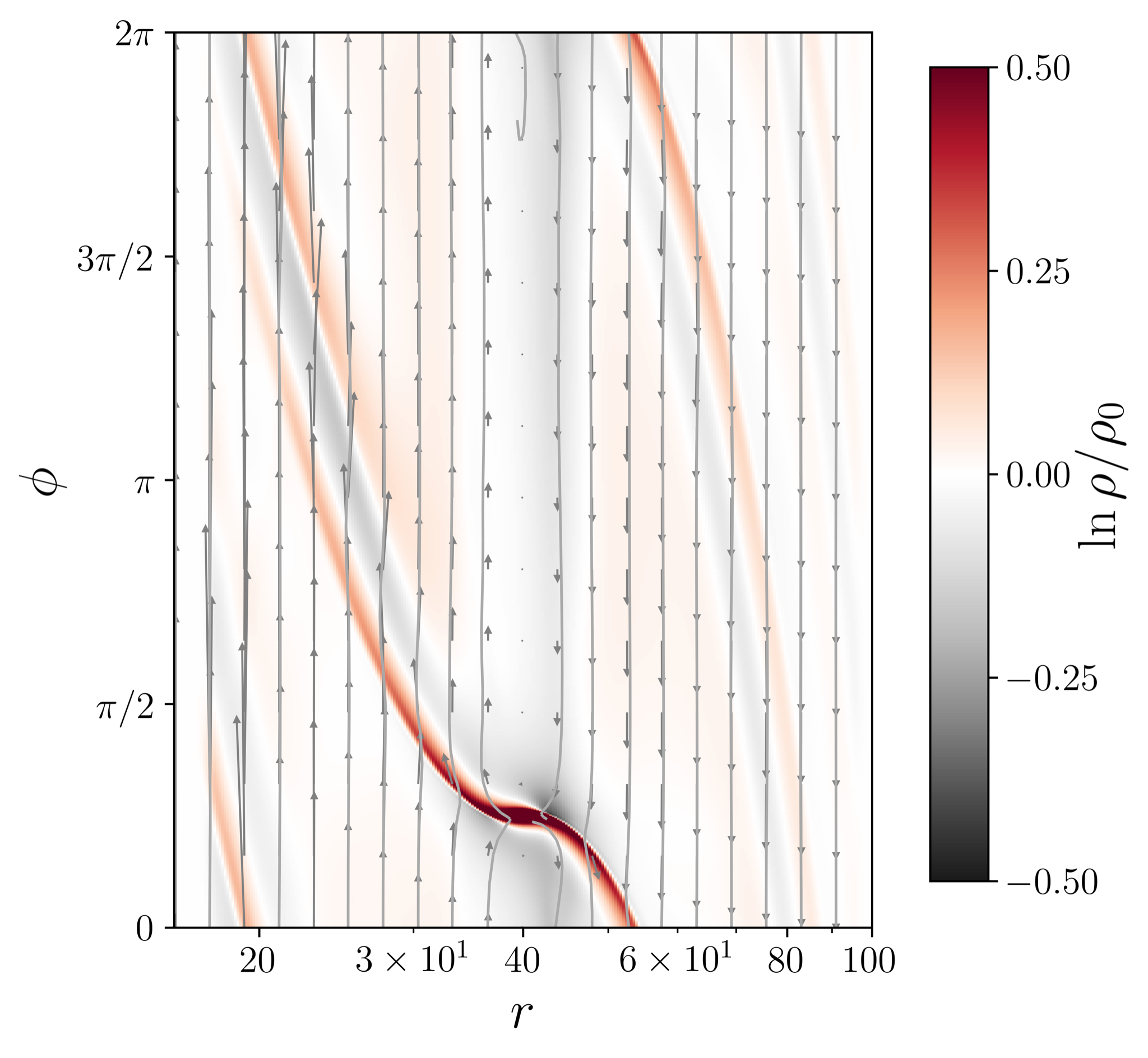}
    \caption{Midplane flow pattern in our disk, in the co-rotating frame of the planet. Streamlines proceed from bottom to top inside the planet radius $r_p = 40 {\rm au}$, and from top to bottom outside the planet radius.}
    \label{fig:flow_midplane}
\end{figure}

\begin{figure}
    \centering
    \includegraphics[width=0.45\textwidth]{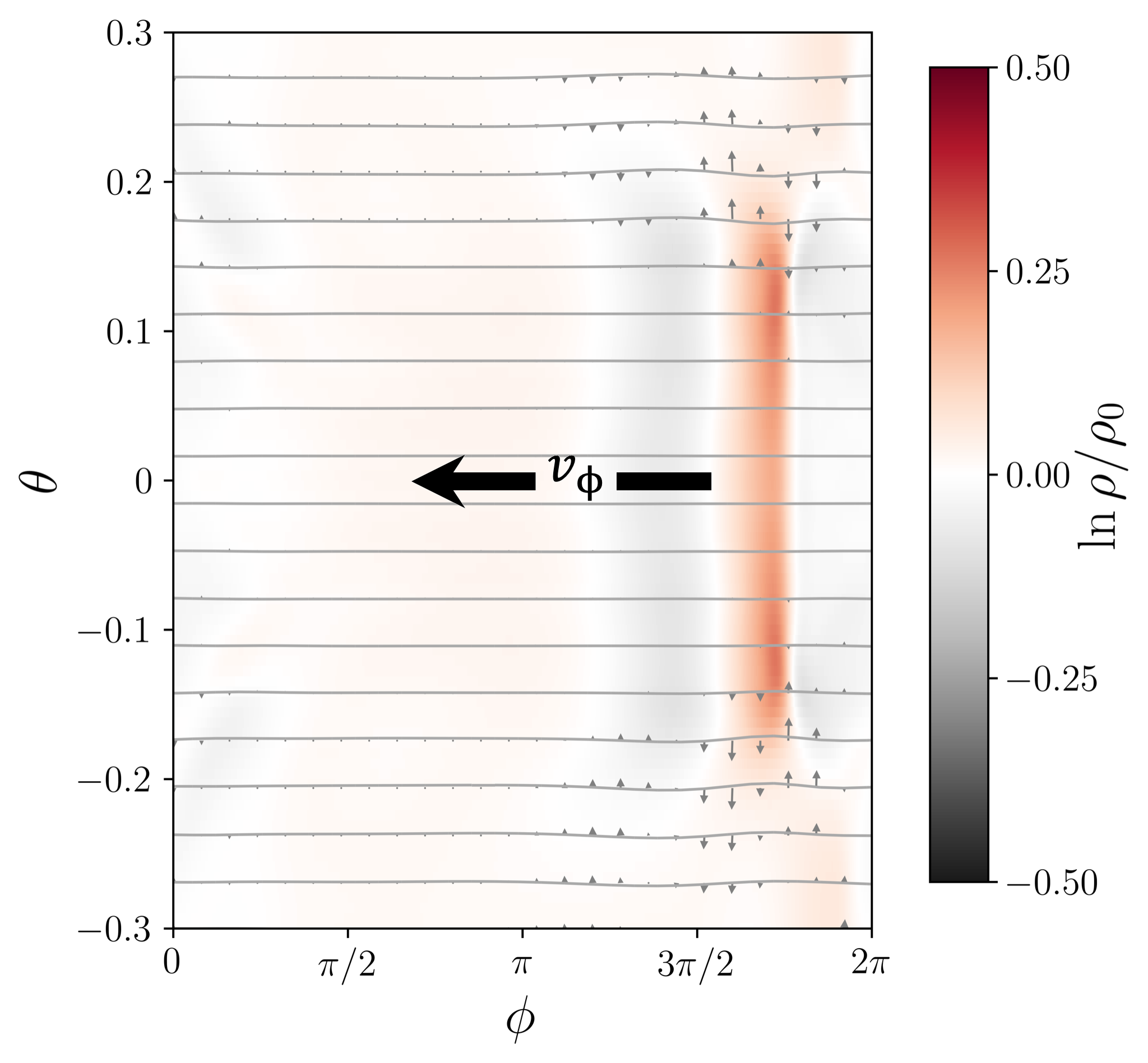}
    \caption{Flow pattern at $r = 1.5 r_p = 60 {\rm au}$, in an azimuthal cut of the disk. Streamlines flow right to left; unlike in Figure \ref{fig:flow_midplane}, vector arrows represent velocity differences from the local initial (quasi-)Keplerian value, rather than from the planet's Keplerian speed.}
    \label{fig:flow_vert}
\end{figure}

\begin{figure}
    \centering
    \includegraphics[width=0.45\textwidth]{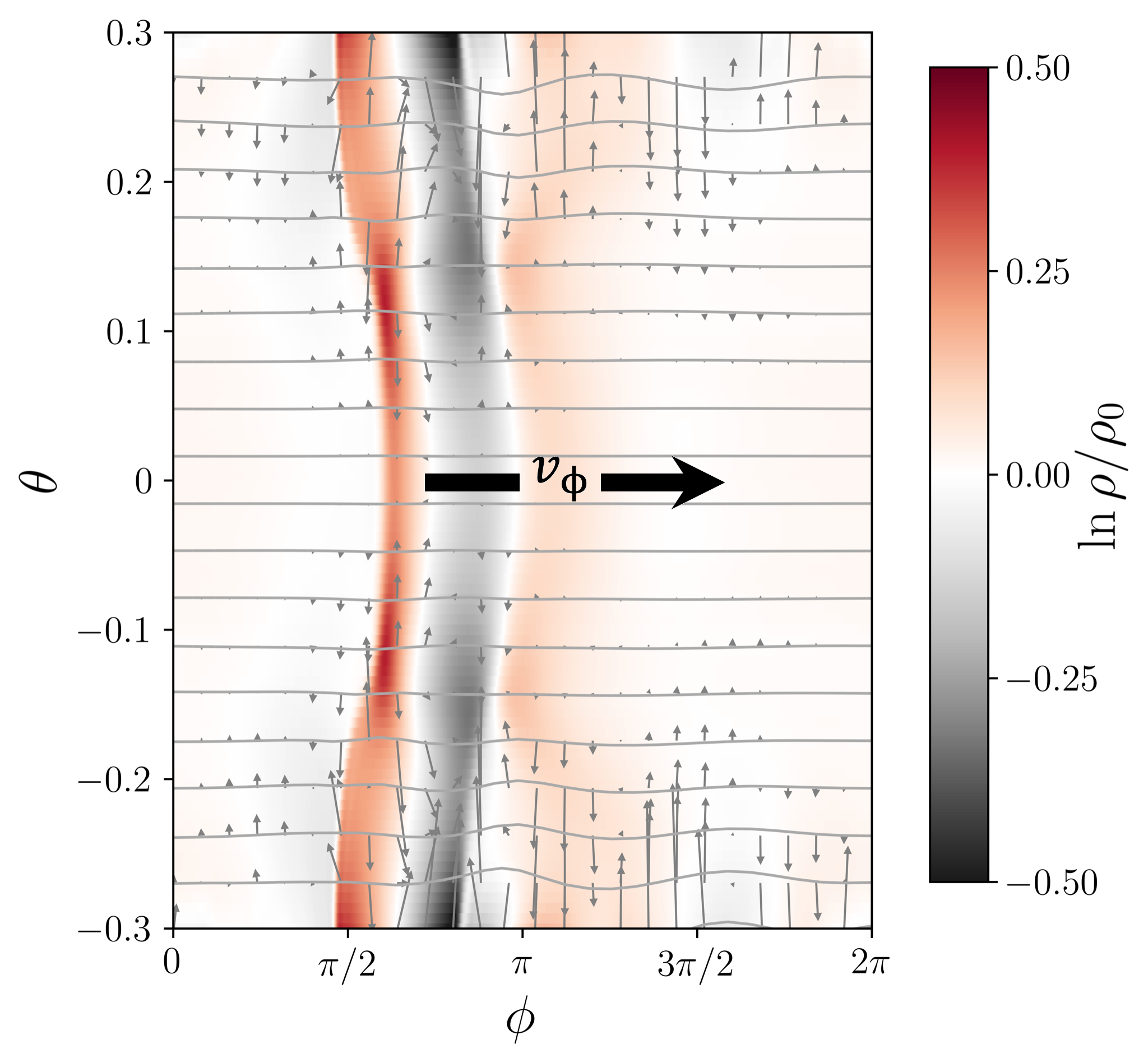}
    \caption{Flow pattern at $r = 0.66 r_p = 26.6 {\rm au}$, in an azimuthal cut of the disk. Streamlines flow left to right; as Figure \ref{fig:flow_vert}, vector arrows represent velocity differences from the local initial (quasi-)Keplerian value.}
    \label{fig:flow_vert_inner}
\end{figure}

\subsection{Flow analysis}\label{sec:flow_analysis}
We next turn our attention to the flow patterns in the disk. In Figures \ref{fig:flow_midplane}, \ref{fig:flow_vert}, \ref{fig:flow_vert_inner}, we plot streamlines of the fluid flow in various disk cuts, along with quivers indicating the instantaneous flow direction (see captions for more details); in the background, we include the density perturbation for reference. Close to the midplane, velocities are essentially restricted to the $r-\phi$ direction as in 2D simulations; however, at the temperature transition, the spiral pattern weakens and bends, with the velocity acquiring a $\theta$-component \citep[see also e.g,][for a more general discussion of vertical spiral velocities]{Boley2006}.

What gives rise to these flow patterns, and in turn, how do they influence the observed spiral perturbations in density, temperature, and kinematics? Our system begins in a background state $\rho_0, T_{g, 0}, \vec{v}_0$, where $\vec{v}_0 = v_{\phi,0} \hat{\vec{\phi}}$ is a divergenceless, axisymmetric, quasi-Keplerian flow set by the stellar potential $\Phi_{*}$ and the initial pressure profile $P_0$. All quantities $(\rho, T_g, T_d, v_r, v_{\theta}, v_{\phi})$ can be expressed in terms of the initial condition and a perturbation, e.g., $\rho = \rho_0 + \rho'$, both of which are in general dependent on space. Working in the rotating frame of the planet, and given that we do not include gas opacities, 
we can write the evolution of the perturbations as follows:
\begin{subequations}
\begin{equation}\label{eq:density_perturb}
    \frac{\partial \rho'}{\partial t} = -\left[\vec{v} \cdot \nabla \rho' - \vec{v}' \cdot \nabla \rho_0 \right] - \left[\rho \nabla \cdot \vec{v}'
    \right]
\end{equation}
\begin{equation}\label{eq:tg_perturb}
\begin{split}
    \frac{\partial T_g'}{\partial t} = -\left[\vec{v} \cdot \nabla T_g' + \vec{v}' \cdot \nabla T_{g,0}\right] - (\gamma - 1) \left[T_g \nabla \cdot \vec{v}' 
    \right] - t_{\rm c}^{-1}\left[T_{g}' - T_{d}'\right]
\end{split}
\end{equation}
\begin{equation}\label{eq:momentum_perturb}
    \frac{\partial \vec{v}'}{\partial t} = -\left[\vec{v} \cdot \nabla \vec{v}' - \vec{v}' \cdot \nabla \vec{v}_0\right] - \left[\nabla P/\rho - \nabla P_0/\rho_0 + \nabla \Phi_p\right] - 2 \vec{\Omega}_p \times \vec{v}'
\end{equation}
\end{subequations}
In the above equations, the partial time-derivative terms represent overall evolution of the quantity at a fixed location; given that the spiral is well-developed at $t = 2500$ yr, and that its pattern speed equals the frame rotation speed, the system is approximately in steady state and these terms net to a small number. The terms in the first set of square brackets, including velocities projected along gradients of various quantities, represent advection of the flow. We consider both the vertical (formally, along the $\hat{\theta}$-direction) and in-plane (formally, along the $\hat{r}$ and $\hat{\theta}$ directions) transport of perturbed density, temperature, and velocity within our stratified disk; we write explicit expressions for these terms in Appendix \ref{sec:adv_terms}.

In the second set of square brackets are source terms representing gas compressibility, $p dV$ work, and pressure-gravity balance in Equations \ref{eq:density_perturb}, \ref{eq:tg_perturb}, and \ref{eq:momentum_perturb} respectively. For simplicity, we aggregate the last of these quantities into

\begin{equation}
    \left(\frac{\partial \vec{v}'}{\partial t}\right)_{\rm source} = -\left[\nabla p/\rho - \nabla p_0/\rho_0 + \nabla \Phi_p\right]
\end{equation}

The final bracketed term in Equation \ref{eq:tg_perturb} represents the exchange of energy between gas and dust. The term $2 \vec{\Omega}_p \times \vec{v}'$ in Equation \ref{eq:momentum_perturb} is the Coriolis acceleration arising from frame rotation. In the second set of square brackets are the source terms, involving divergences of velocities for the scalar quantities, and gradients of pressure/gravitational potential for the velocity. 

\begin{figure}
    \centering
    \includegraphics[width=0.45\textwidth]{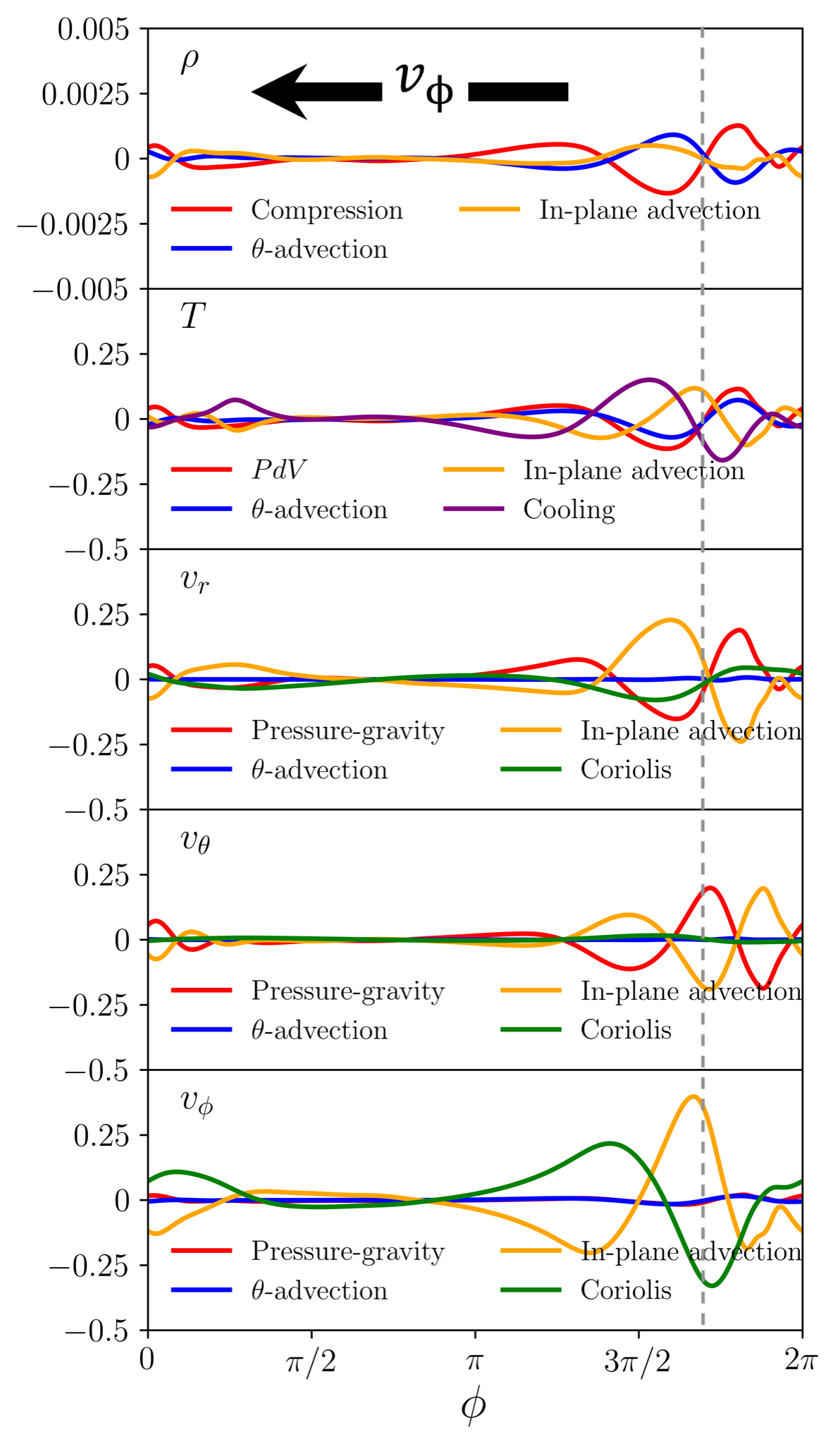}
    \caption{Source (red) and advective (blue) terms at a fixed $r = 1.5 r_p$ and $\theta = 0.2$, plotted over azimuthal angle $\phi$. From top to bottom, we plot these terms for $\rho$ (in units of $\Omega_p \rho_{0, p}$), $T_g$ ($\Omega_p T_{g, 0, p}$), $v_r$, $v_{\theta}$, and $v_{\phi}$ ($\Omega_p c_{s, \rm iso, p}$), where the $p$ subscript indicates quantities taken at the planet location in the initial condition. A thin, grey, dashed line passes through the azimuthal peak of the density spiral, showing the significant offset between terms driving spiral perturbations in each quantity.
    }
    \label{fig:source_transport_outer}
\end{figure}

We plot all of these terms and vertical-advective terms in a cut at $r = 1.5 r_p$ and $\theta = 0.2$ in Figure \ref{fig:source_transport_outer}. As fluid in this upper-disk region enters the spiral density wave, its quasi-Keplerian orbit is perturbed down ($v_{\theta}$ pressure-gravity) toward the midplane---against $\nabla \rho_0$, but along $\nabla T_{g,0}$---decreasing the local $T_g$ while increasing $\rho$ ($\theta$-advection). Because this flow pattern (which also includes a perturbation in $v_r$) has a nonzero divergence, it also increases both $\rho$ and $T_g$ by compression and $p dV$ work, respectively. These terms are balanced by quasi-Keplerian transport of gas through the spiral pattern (``in-plane'' advection), and additionally for $T_g$, collisional relaxation to the background dust temperature (cooling), which hold the system in an approximately steady state.

For velocity, vertical gradients are much weaker, so $\theta$-velocity does not play a significant role. Instead, perturbations in the $v_r$ and $v_{\theta}$ components are governed primarily by the planet-driven density wave (pressure-gravity) and counterbalanced by in-plane transport across the spiral. Along the $\phi$-component, the pressure gradient is weak, and $v_{\phi}$ is governed instead by a balance between in-plane advection and Coriolis terms. The fact that $v_r$ and $v_{\rm phi}$ are somewhat out of balance reflects the fact that over the long term, planet-driven spirals open a gap in the disk.

\subsection{Equilibrium temperature}\label{sec:equilibrium_temp}
Disk-planet interaction not only heats the disk locally through $p dV$ work, but non-locally by changing the background radiation field. Spiral arms, for instance, push disk material to higher altitudes, intercepting direct stellar irradiation while gently shadowing the regions behind them. Closer to the midplane, the accumulation of material in circumplanetary regions leads to more clear, radially-directed shadowing. The gas heating in both regions, whether through spiral compression or gas accretion, also makes them weak \footnote{The circumplanetary region becomes a much stronger source when accretion luminosity is accounted for; see Section \ref{sec:acclum_analysis} for more details.} sources of radiation. All this impacts the equilibrium temperature $T_{eq}$, defined by the equation

\begin{equation}
    0 = S_d^{\rm irr} + \rho_d \kappa_d(T_{eq}) (E_r - a_r T_{eq}^4) \,
\end{equation}

which incorporates both stellar irradiation and the thermalized radiation field, and whose solution we find using Newton-Raphson iterations. Because these effects depend on the (inherently non-local) transport of radiation, they cannot be accounted for using a $\beta$-cooling approach; conversely, the deviation of $T_{d, eq}$ from the initial condition $T_{d, 0}$ provides a good measure of the suitability of a $\beta$-cooling prescription for a particular problem.

\begin{figure}
    \centering
    \includegraphics[width=0.45\textwidth]{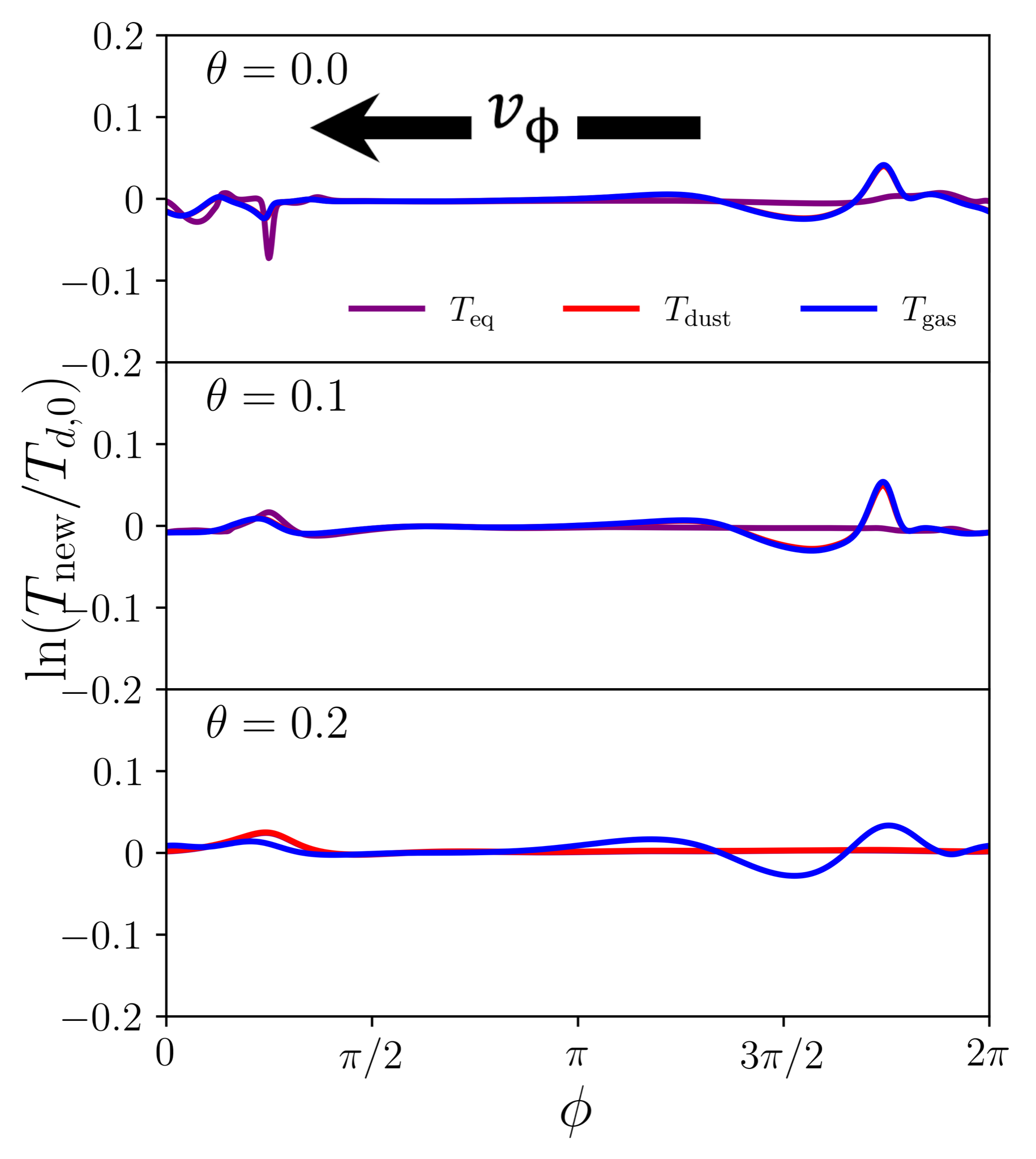}
    \caption{Plots of background equilibrium, gas, and dust temperatures in our 3T, Saturn-mass simulations, at selected altitudes above the midplane for a fixed $r = 1.5 r_p$. Background temperatures deviate by only a few percent from the initial condition, most strongly in the region of the planetary shadow ($\phi \approx \pi/4$) and somewhat more weakly near the Lindblad spiral. $T_d$ agrees well with $T_g$ at $\theta = 0.0$ and $0.1$ (and is covered by the line for $T_g$), and with $T_{\rm eq}$ at $\theta = 0.2$ (covering the line for $T_{\rm eq}$).}
    \label{fig:equilibrium_temp}
\end{figure}

In Figure \ref{fig:equilibrium_temp}, we plot deviations of $T_{eq}$, $T_{d}$, and $T_{g}$ in our 3T, Saturn-mass simulation; we fix $r = 1.5 r_p$ and $\theta = \{0, 0.1, 0.2\}$, and plot the deviations as a function of azimuthal angle. We find that deviations in equilibrium temperature are strongest at $\phi \approx \pi/4$, the angular position of the planet, resulting from midplane shadowing and upper-atmosphere exposure to radiation. This corresponds to the bright ``pseudo-arm'' observed in Monte Carlo radiative-transfer (MCRT) modeling of near-infrared scattered light from simulated planetary spirals \citep{Muley2021}. The deviations in $T_{\rm eq}$ located approximately $\pi/4$ rad from the spiral arms are much weaker; given that they are not centered on the gas temperature bump, we attribute them to rearrangement of disk density affecting transport of stellar and reprocessed radiation, rather than emission from the spiral itself found by \cite{Ziampras2023}.

As in \cite{Muley2023}, we find that dust and gas temperatures largely agree with each other at lower disk layers, whereas in the upper atmosphere, longer dust-gas coupling mean that the gas temperature reflects $p dV$ work from the spirals, while the dust temperature closely tracks the equilibrium temperature. The generally small deviation of $T_{\rm eq}$ explains why the $\beta$-cooling approach discussed in Section \ref{sec:beta_cooling} provides generally accurate results for non-accreting, Saturn-mass planets. However, for other setups---such as those we discuss in the following sections---this need not be the case, and radiation hydrodynamics are essential to obtaining physically consistent results.

\section{Parameter study}\label{sec:parameter_study}

In what follows, we test three different planet masses ($M_p = \{5 \times 10^{-5}, 3 \times 10^{-4}, 1 \times 10^{-3} M_{\odot}$, corresponding to Neptune-, Saturn-, and Jupiter-mass, respectively)\footnote{Given the scale height ratio $h_p = 0.0646$ at the planet location, these correspond to thermal masses $q_{\rm th}$ of 0.18, 1.12, and 3.74 respectively, spanning the sub-thermal, thermal, and super-thermal mass regimes.} with no accretion luminosity, and two planetary accretion luminosities ($L_{\rm acc, p} = \{0, 1 \times 10^{-3}\}$) with a Saturn-mass planet. In order to capture changes to the background radiation field---which become especially pronounced for high-mass or accreting protoplanets---we use the full three-temperature scheme for all of these simulations.

\begin{figure*}
    \centering
    \includegraphics[width=0.45\textwidth]{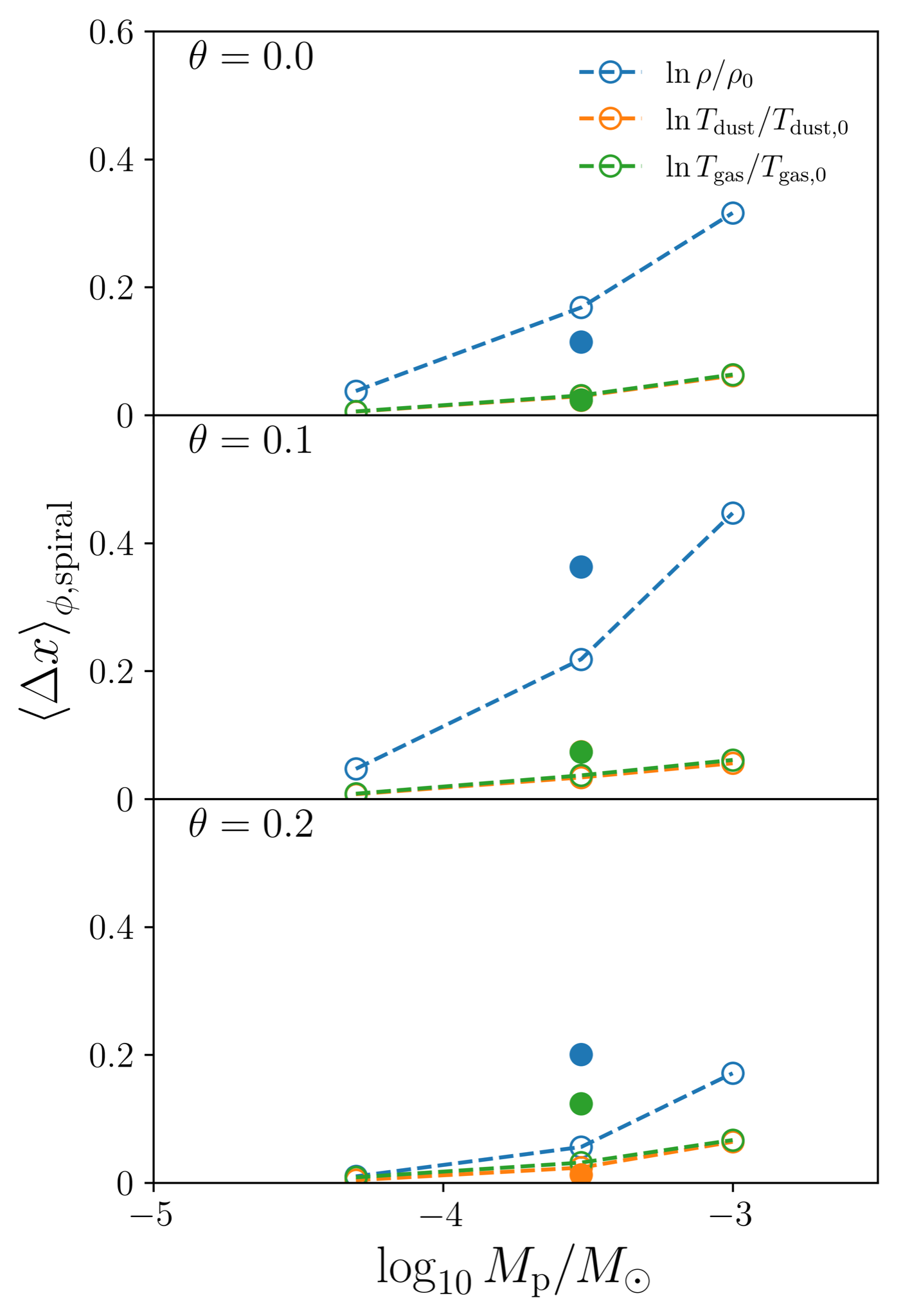}
    \includegraphics[width=0.45\textwidth]{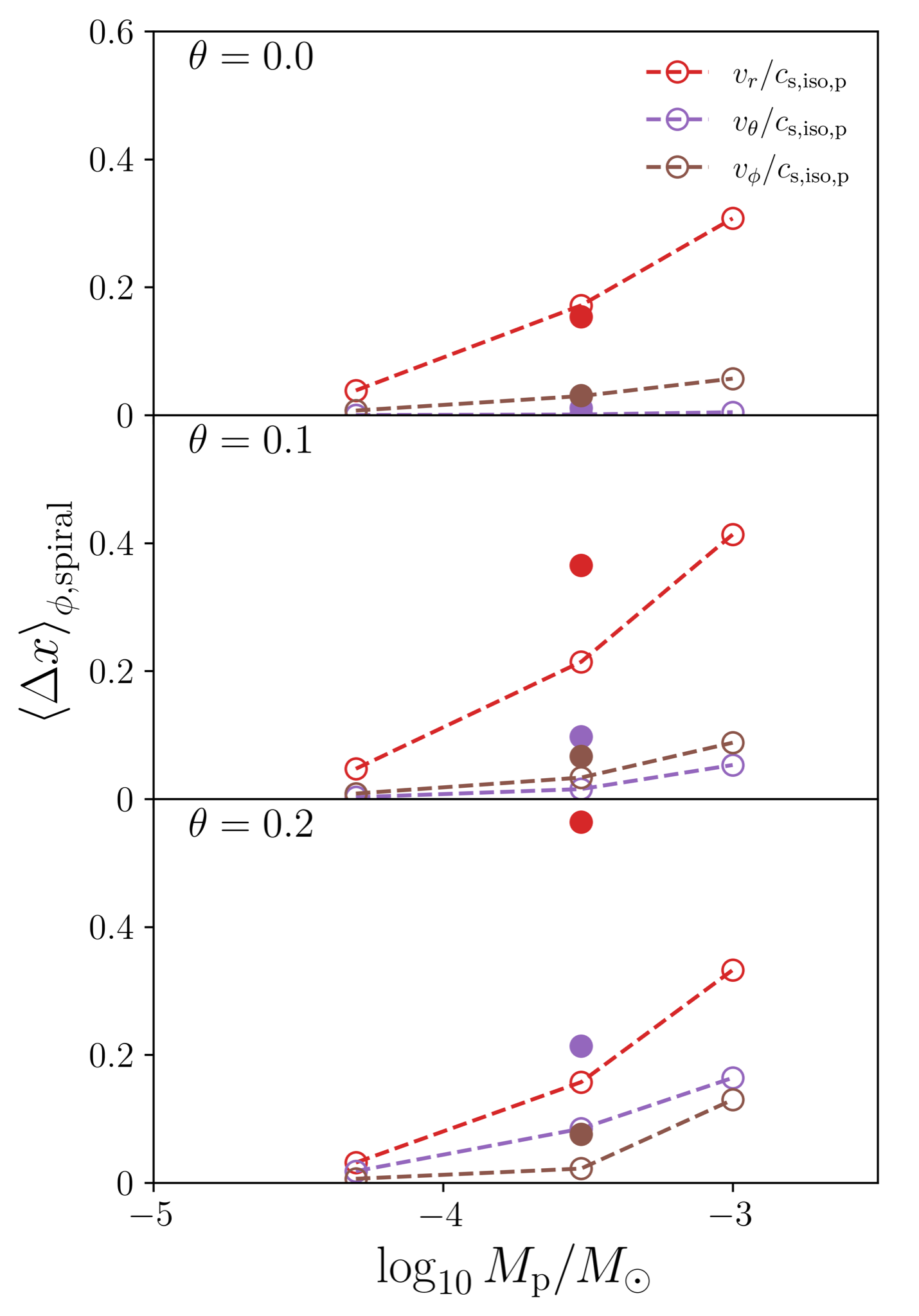}
    \caption{Measurement of the azimuthal perturbation in each normalized quantity $x$, averaged over an azimuthal range of $\phi_{\rm peak} \pm 2 h_p$ according to Equation \ref{eq:spiral_avg}. Densities and temperatures are plotted in the left panel while velocities are in the right. Open circles indicate non-accreting planets while closed circles indicate those with accretion luminosity (see Section \ref{sec:acclum_analysis} for discussion). At $z/r \approx \theta = 0.2$, where dust and gas are not well-coupled, the plotted $T_{\rm dust}$ amplitude reflects the radial ``pseudo-arm'' (see discussion in Section \ref{sec:equilibrium_temp}) rather than the Lindblad spiral.}
    \label{fig:spiral_amplitude_quantitative}
\end{figure*}

\begin{figure*}
    \centering
    \includegraphics[width=0.9\textwidth]{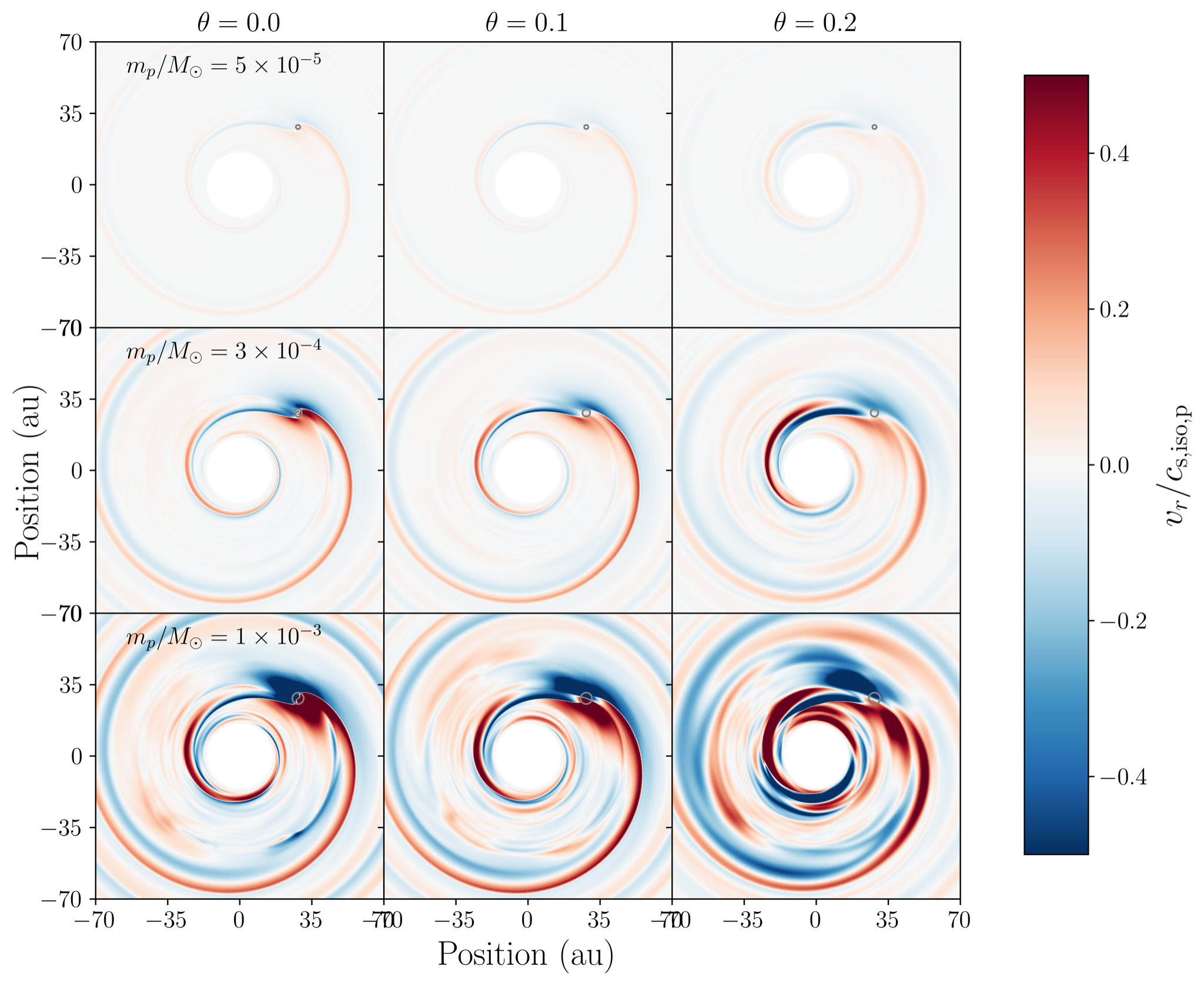}
    \caption{Radial velocities $v_r$ from 3T, zero-accretion-luminosity simulations with Neptune-mass \textit{(top row)}, Saturn-mass \textit{(middle row)}, and Jupiter-mass \textit{(bottom row)} planets, at various altitudes in the disk \textit{(left, middle, right columns)}. Grey circles indicate the planetary Hill radius. All $v_r$ values are expressed as a function of initial isothermal sound speed at the location of the planet, $c_{\rm iso, p} = \sqrt{p_{p,0}/\rho_{p, 0}}$. }
    \label{fig:rad_vel_perturbation}
\end{figure*}

\begin{figure*}
    \centering
    \includegraphics[width=0.9\textwidth]{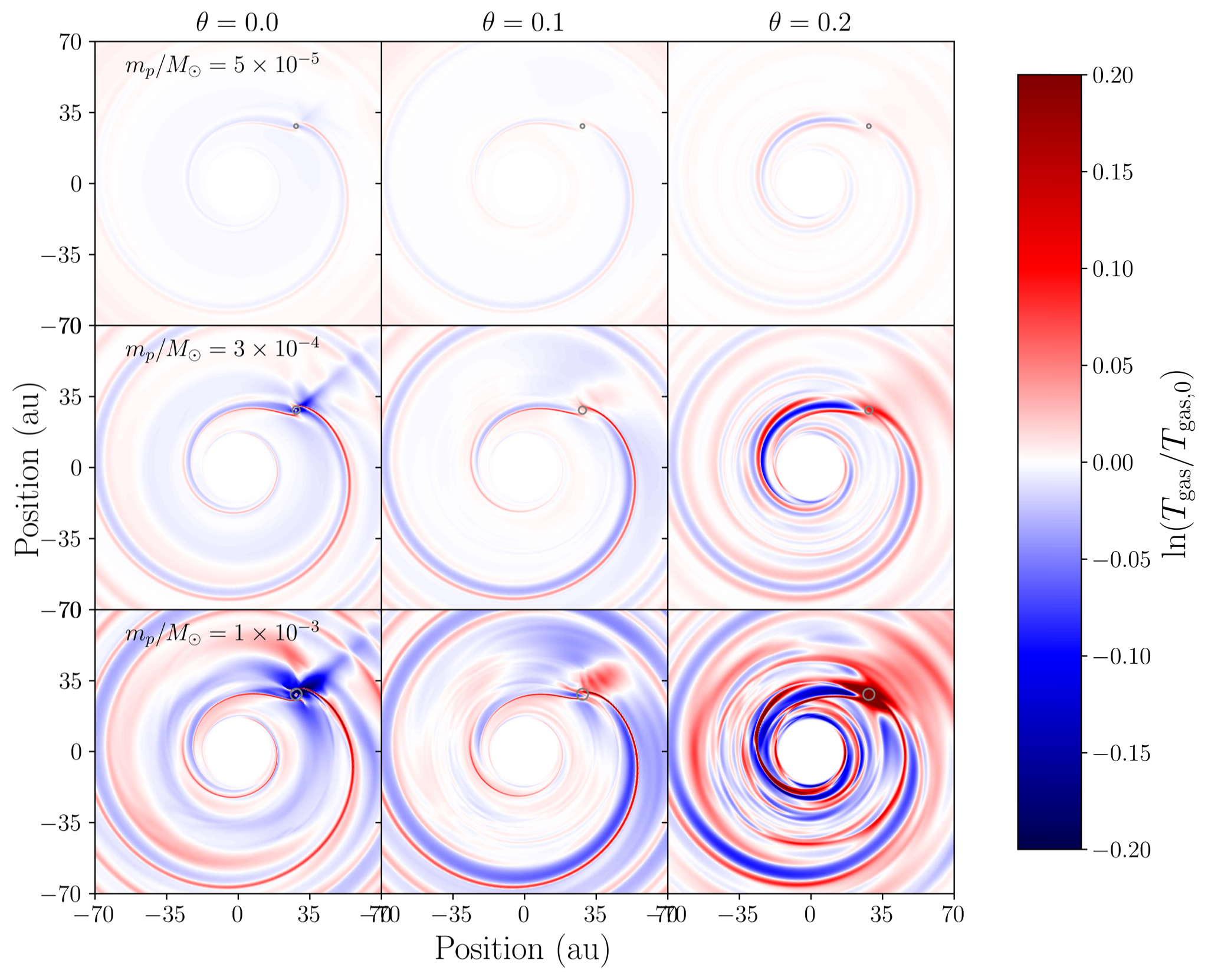}
    \caption{Gas temperature perturbation for Neptune-mass \textit{(top row)}, Saturn-mass \textit{(middle row)}, and Jupiter-mass \textit{(bottom row)} planets, at various altitudes in the disk \textit{(left, middle, right columns)}. Grey circles indicate the planetary Hill radius. All temperature values are expressed with respect to the initial condition $T_{\rm g, 0}$. }
    \label{fig:temp_perturbation}
\end{figure*}

\subsection{Planet mass}\label{sec:planet_mass}

In Figure \ref{fig:spiral_amplitude_quantitative}, we plot the spiral-averaged perturbation amplitude $\left<\Delta x\right>_{\phi, \rm spiral}$---where $x$ can be any one of various normalized quantities $(\ln \rho/\rho_0, \ln T_g/T_{g,0}, \ln T_d/T_{d,0}, v_r/c_{\rm s, iso, p}, v_{\theta}/c_{\rm s, iso, p}, v_{\phi}/c_{\rm s, iso, p})$, and $\Delta x$ its deviation from the initial condition---as a function of planet mass. As in previous figures, we fix $r = 1.5 r_p$ and test $\theta = \{0.0, 0.1, 0.2\}$ above the midplane. We define
\begin{equation}\label{eq:spiral_avg}
    \left<\Delta x\right>_{\phi, \rm spiral}(r, \theta) = \frac{1}{4h_p}\int_{\phi_{\rm peak} - 2 h_p}^{\phi_{\rm peak} + 2 h_p} \Delta x (r, \theta, \phi) d\phi
\end{equation}
which is analogous to the definition in \cite{Muley2021}, with the important distinction that in the present work the amplitude is not vertically averaged. Open circles indicate simulations without accretion luminosity, while filled circles correspond to our simulation with it (Section \ref{sec:acclum_analysis}).

As found in previous works \citep{Fung2015,Dong2017,Muley2021}, spiral amplitude increases substantially with planet mass, irrespective of the measure used. At $\theta = 0.0$ and $\theta = 0.1$, $\rho$ and $v_r/c_{\rm s, iso, p}$ perturbations follow each other closely; both weaken in the upper atmosphere, but the density perturbation much more so. $v_\theta/c_{\rm s, iso, p}$ is nearly zero in the disk midplane, but rises with increasing altitude (see Section \ref{sec:flow_analysis}). $v_{\phi}/c_{\rm s, iso, p}$ is relatively weak in most cases, but is stronger for high-mass cases at $\theta = 0.1, 0.2$ thanks to the distortions introduced by the circumplanetary flow. At every altitude and planet mass, temperature deviations tend to be weaker than those in density and velocity. At $\theta = 0.0, 0.1$ where dust and gas are well-coupled, they reflect both the Lindblad spiral and the radial ``pseudo-arm'' instead (Section \ref{sec:equilibrium_temp}), but at $\theta = 0.2$, gas temperature primarily reflects the Lindblad spiral while dust temperature reflects the pseudo-arm.

For a more qualitative understanding, we show 2D polar cuts of $v_r$ in Figure \ref{fig:rad_vel_perturbation}. Sub-thermal, Neptune-mass planets excite well-behaved kinematic spirals clearly distinct from the laminar background, however, these spirals are rather weak, with typical $|v_r| \lesssim 0.1 c_{\rm s, iso, p}$---corresponding to a physical velocity $\lesssim 30 \si{\meter\per\second}$ in our disk model. In the Saturn-mass case, velocity spirals have a similar shape, but are significantly stronger---by a factor of ${\sim}4$ in the outer disk (Figure \ref{fig:spiral_amplitude_quantitative}), and even more so in the inner disk. Here, flows at the corotation radius and in the vicinity of the planet, which can be identified with the velocity kinks observed in channel maps \citep[e.g.,][]{Pinte2019}, become significantly more prominent.

In the Jupiter-mass case, Lindblad spirals become somewhat stronger, with higher-order spirals \citep{Bae2018a,Bae2018b} particularly visible in the inner and upper disks. However, the observed $v_r$ signature is dominated by flows near the planet, which become larger and faster thanks to the greater planet mass and gravitational sphere of influence. In the upper atmosphere, this flow takes on a wing-like shape, and acts to bring material directly above/below the planet, where is funneled toward the planet's Hill radius through the poles \citep[e.g.,][]{Fung2015b,Fung2019}. At all altitudes, the background flow also becomes more turbulent and unsteady.

To complement our understanding from the kinematic signatures, we present perturbations in the gas temperature $T_g$ in Figure \ref{sec:equilibrium_temp}, taken at the same 2D polar cuts as in Figure \ref{fig:rad_vel_perturbation}. In all simulations, we observe a radial shadow from the circumplanetary region (see Section \ref{sec:equilibrium_temp}) whose strength increases with the size of the circumplanetary region and thus, with planet mass. Especially for higher-mass cases, the fact that the integrated one-sided Lindblad torque scales as $\mathcal{T}_{L,+} = C_{L,+} (M_p/M_*)^2 h^{-3} \propto T_g^{-3/2}$ \citep{Tsang2011} may contribute to strengthening the outer spiral. In the upper atmosphere, the funneling of gas into a flow toward the midplane along the planetary pole leads to compressional heating, manifesting as a hot spot in $T_g$ above the planet location.

Thermal relaxation, first by gas-grain collision and then by thermal emission from heated dust grains, dissipates $p dV$ work to the radiation field. For the disk setup we choose, the effective relaxation timescale ($\beta = 0.1-1$) is approximately equal to or less than the spiral crossing time. As a result, the $T_g$ spiral has a multi-band structure, reflecting the initial compression when the gas strikes the spiral (high-temperature band), expands behind the spiral (low-temperature band), and compresses again to return to the background density (high-temperature band). The first high-temperature band is dominant in the midplane, but both roughly even in the disk upper atmosphere. This stands in contrast to the effectively-adiabatic situation where thermal relaxation is much slower than spiral crossing \citep{Miranda2020b,Muley2021}, where the $T_g$ spiral reflects the accumulated total of compression and expansion rather than individual short phases of it. This picture is clear up to Saturn-mass, but less so at Jupiter-mass, where the $T_g$ Lindblad spiral is distorted by the effects of circumplanetary and turbulent flows.

\begin{figure*}
    \centering
    \includegraphics[width=0.9\textwidth]{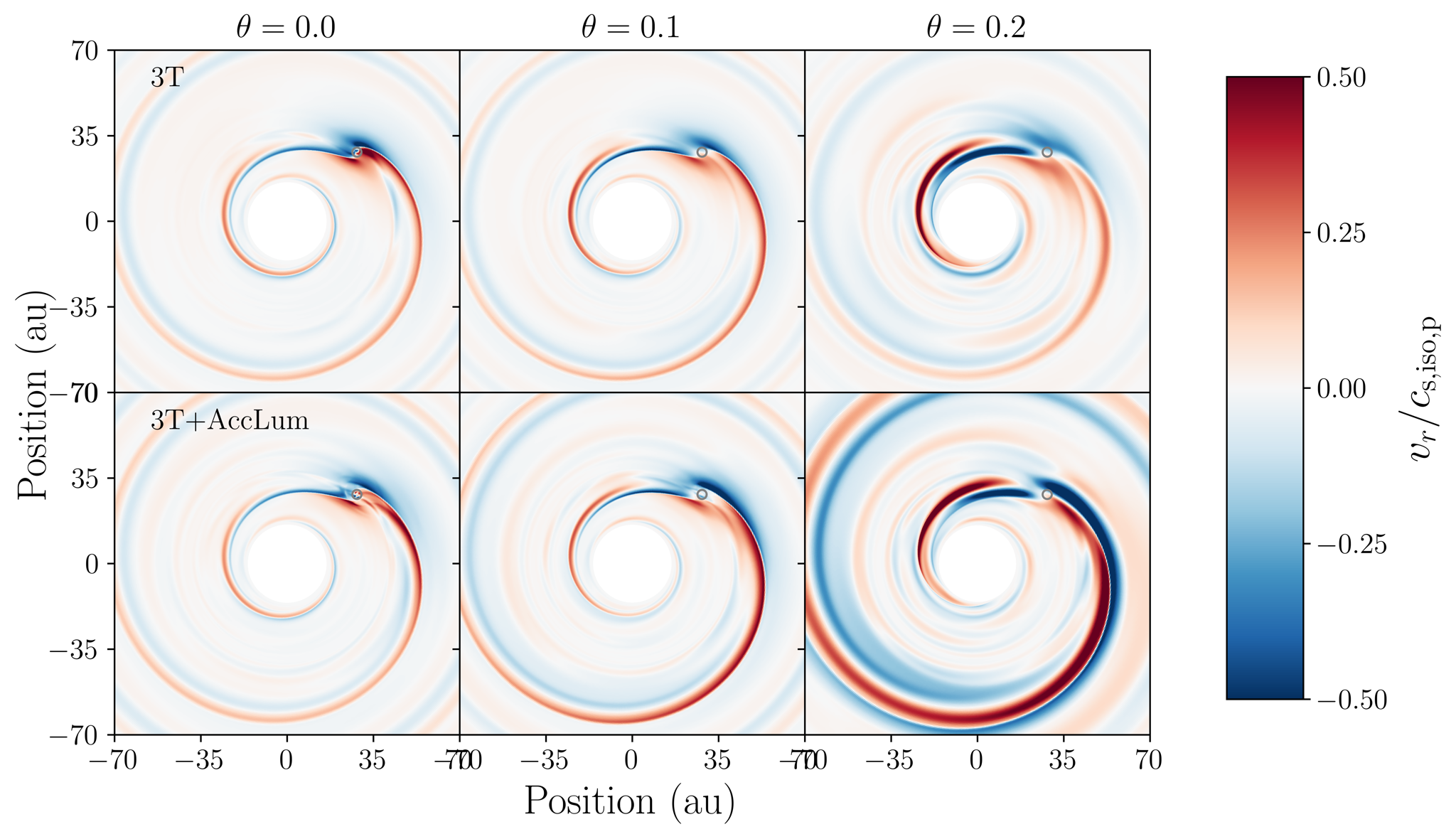}    
    \includegraphics[width=0.9\textwidth]{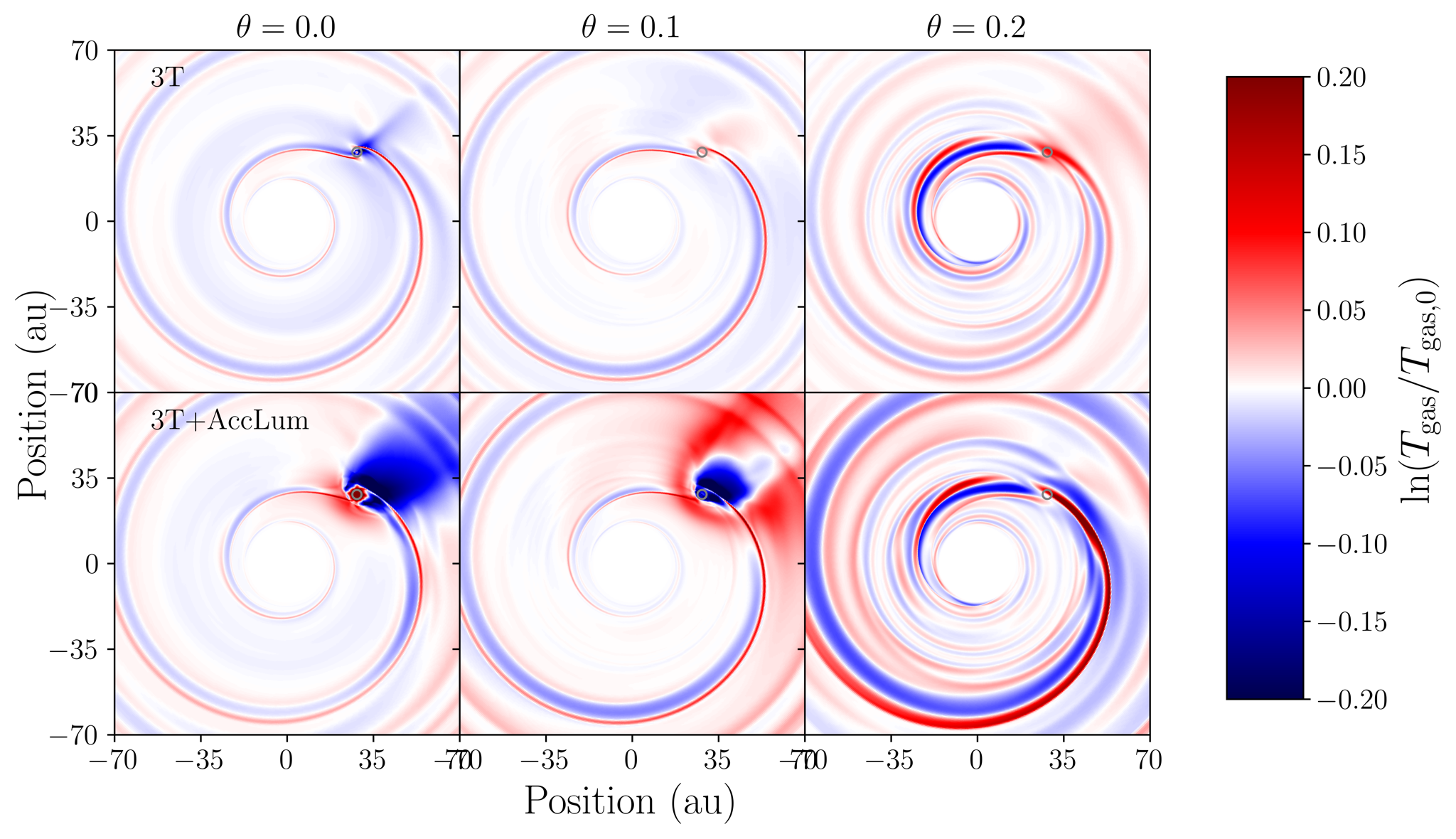}
    \caption{Difference from the initial condition in $v_r$ (\textit{above}) and $T_g$ (\textit{below}), for three-temperature simulations of Saturn-mass planets without and with accretion luminosity. The strongly luminous planet alters the vertical and azimuthal structure of the circumplanetary region, causing strong shadows behind the planet and greatly enhancing the kinematic signatures of accretion in the outer disk.}
    \label{fig:acc_lum_changes}
\end{figure*}

\begin{figure}
    \centering
    \includegraphics[width=0.45\textwidth]{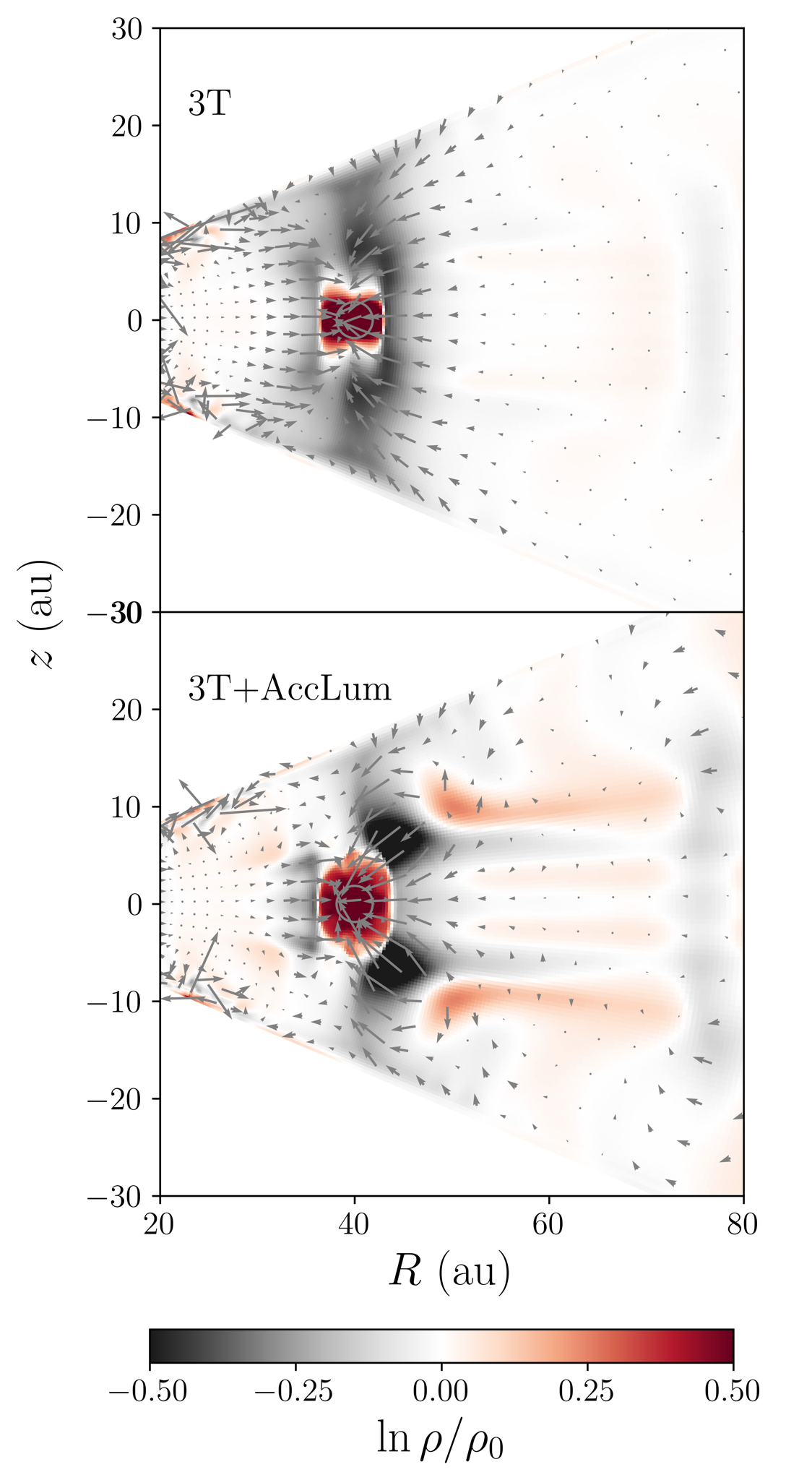}
    \caption{Plot of density perturbation in a cut through $\phi_p = \pi/4$, as a function of cylindrical radius $R$ and vertical position $z$, for 3T simulations of Saturn-mass planets without (\textit{above}) and with (\textit{below}) accretion luminosity. Grey arrows indicate the velocity field, while the grey circle encloses the planetary Hill radius. When accretion luminosity is included, the circumplanetary envelope becomes larger and the flow around it is altered.}
    \label{fig:acclum_dens_cyl}
\end{figure}

\subsection{Accretion luminosity}\label{sec:acclum_analysis}
In recent years, the effect of planetary accretion luminosity has been studied in parametrized 2D global disk simulations \citep{Montesinos2015, Montesinos2021, Garate2021}, and in local, 3D simulations with full radiative transfer \citep{Szulagyi2017} that emphasize the circumplanetary disk. We add to this body of work with our 3D global simulations, including three-temperature radiation transport and a luminous planet. The $L_{\rm acc, p} = 10^{-3} L_{\odot}$ we use corresponds to a mass accretion rate of of $\dot{M} = 7 M_J/{\rm Myr}$, given the fiducial planet mass $M_p = 4 \times 10^{-4} M_{\odot}$ and assuming a typical radius of $2 R_J$ for the forming planet. For this mass, such an accretion rate can only be sustained over brief periods. We thus believe that this scenario, along with the no-accretion base-case, bracket the range of accretion luminosities that the planet might experience during its growth.

We revisit Figure \ref{fig:spiral_amplitude_quantitative}, in which spiral amplitudes from our simulations with accretion luminosity are plotted with filled circles. At $r = 1.5 r_p$ and $\theta = 0$, accretion luminosity somewhat weakens the $\rho$ and $v_r$ perturbations, while leaving the others unchanged. At higher altitudes, by contrast, most perturbations are significantly strengthened by accretion luminosity---sometimes by factors of 2 or more---with respect to the non-luminous planet case.

Along the lines of 
Figures \ref{fig:rad_vel_perturbation} and \ref{fig:temp_perturbation}
, we also make qualitative plots of kinematic and temperature perturbations at various disk cuts, and present these in Figure \ref{fig:acc_lum_changes}. With planetary accretion luminosity, the circumplanetary envelope becomes more pressure-supported, with the hot core overflowing the planetary Hill radius. Close to the midplane, this disrupts the orderly Doppler-flip kinematic signature and enhances the Lindblad spirals in both temperature and velocity, without changing their overall morphology. It also significantly strengthens the outer radial shadow \citep{Montesinos2021} through its impact on the envelope's vertical and azimuthal structure.

In the upper atmosphere, however, accretion luminosity intensifies the spiral and changes its shape. The radial velocity shows a double-banded structure---as opposed to the single-banded one expected without accretion luminosity---while the temperature shows a cold-hot-cold band structure, rather than hot-cold-hot. With accretion, the spiral also remains strong much farther away from the planet than without.

To understand these results better, we plot the density perturbation in Figure \ref{fig:acclum_dens_cyl}, at a vertical cut through the planet's azimuth at $\phi_p = \pi/4$. In this view, it is clear that accretion luminosity puffs the circumplanetary region vertically. This changes the flow pattern significantly, with accretion primarily happening by material diagonally striking the edge of the envelope, rather than being funneled downward into a narrow polar flow perpendicular to the disk midplane, as is classically the picture \citep[e.g.,][]{Fung2015b}. This is responsible for many of the observed changes to kinematic and thermal signatures, particularly in the upper atmosphere. Moreover, this means that material flowing into the Hill sphere has higher angular momentum in the accreting than non-accreting case, and as such, preferably gets expelled outward.

\section{Observational implications}
\begin{figure*}
    \centering
    \includegraphics[width=0.9\textwidth]{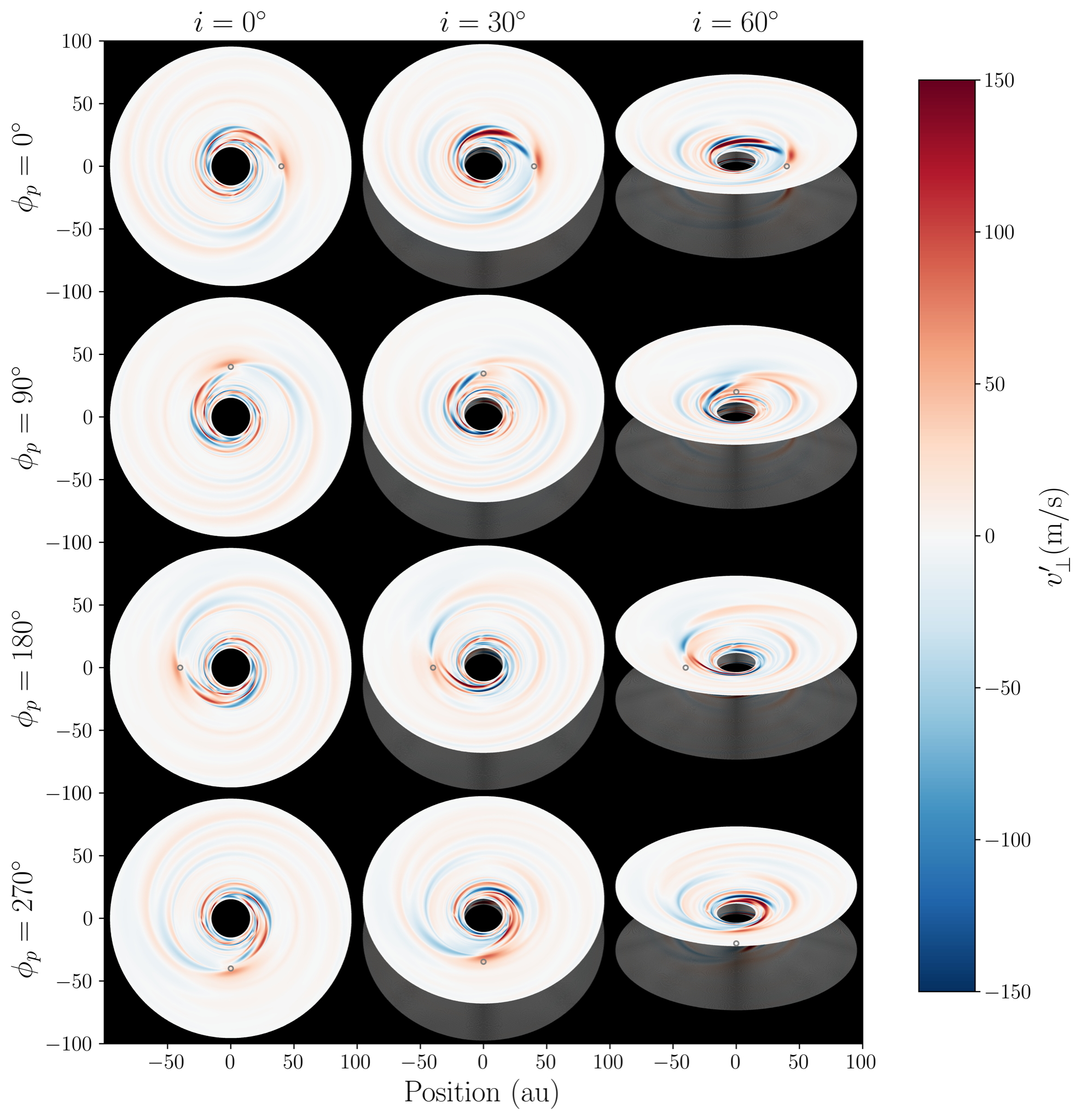}
    \caption{Sky-projected velocity perturbation $v'_\perp$ for our fiducial Saturn-mass planet, in a cut at $\theta = 0.3$ radians above the midplane.}
    \label{fig:3t_spv}
\end{figure*}

\begin{figure*}
    \centering
    \includegraphics[width=0.9\textwidth]{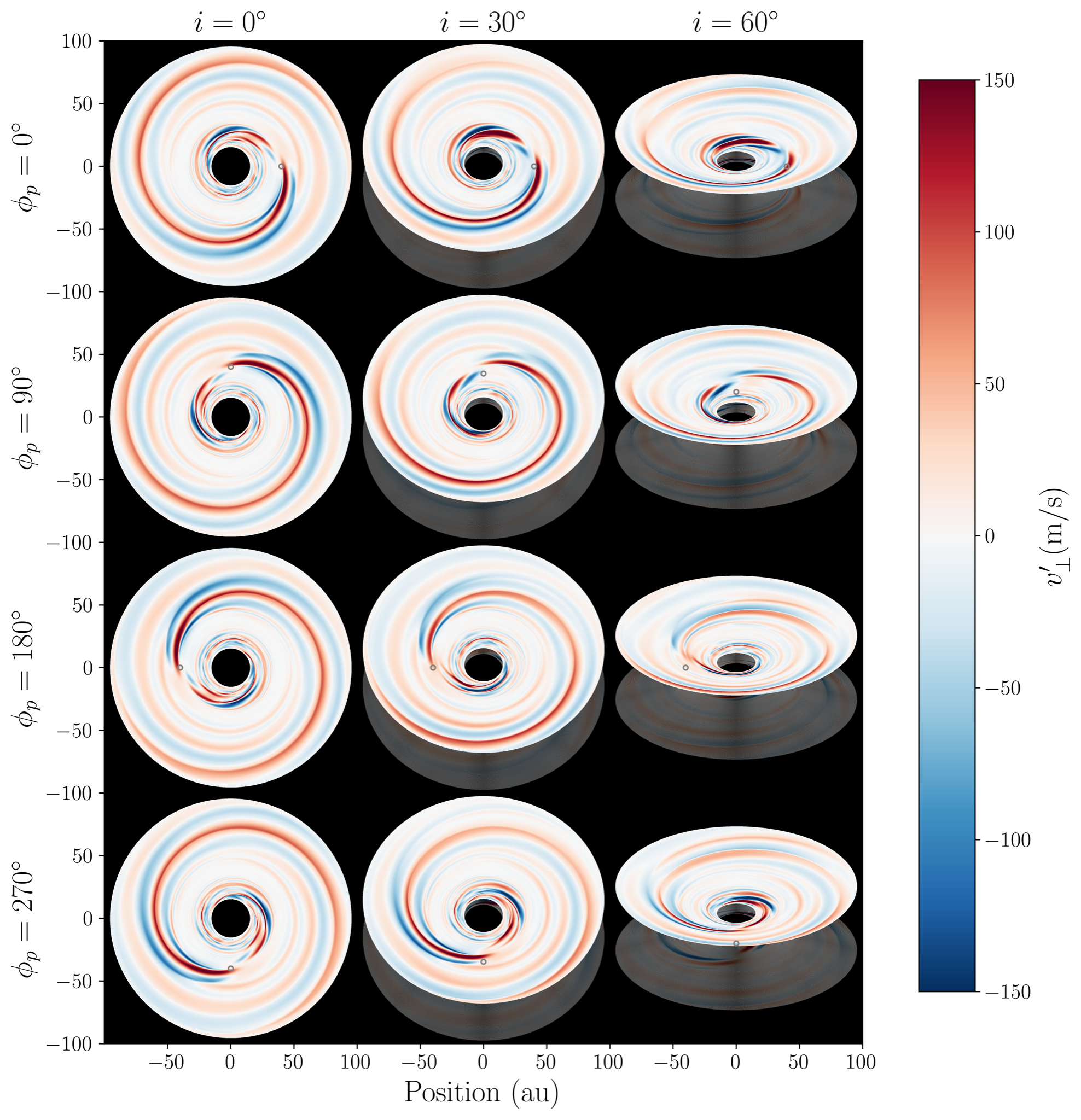}
    \caption{Sky-projected velocity perturbation $v'_\perp$ for a Saturn-mass planet with an accretion luminosity $L_{\rm acc, p} = 10^{-3} L_{\odot}$. Compared to the fiducial case, the outer spiral becomes significantly stronger and extends through a larger radial range of the disk.}
    \label{fig:acclum_spv}
\end{figure*}

\begin{figure*}
    \centering
    \includegraphics[width=0.9\textwidth]{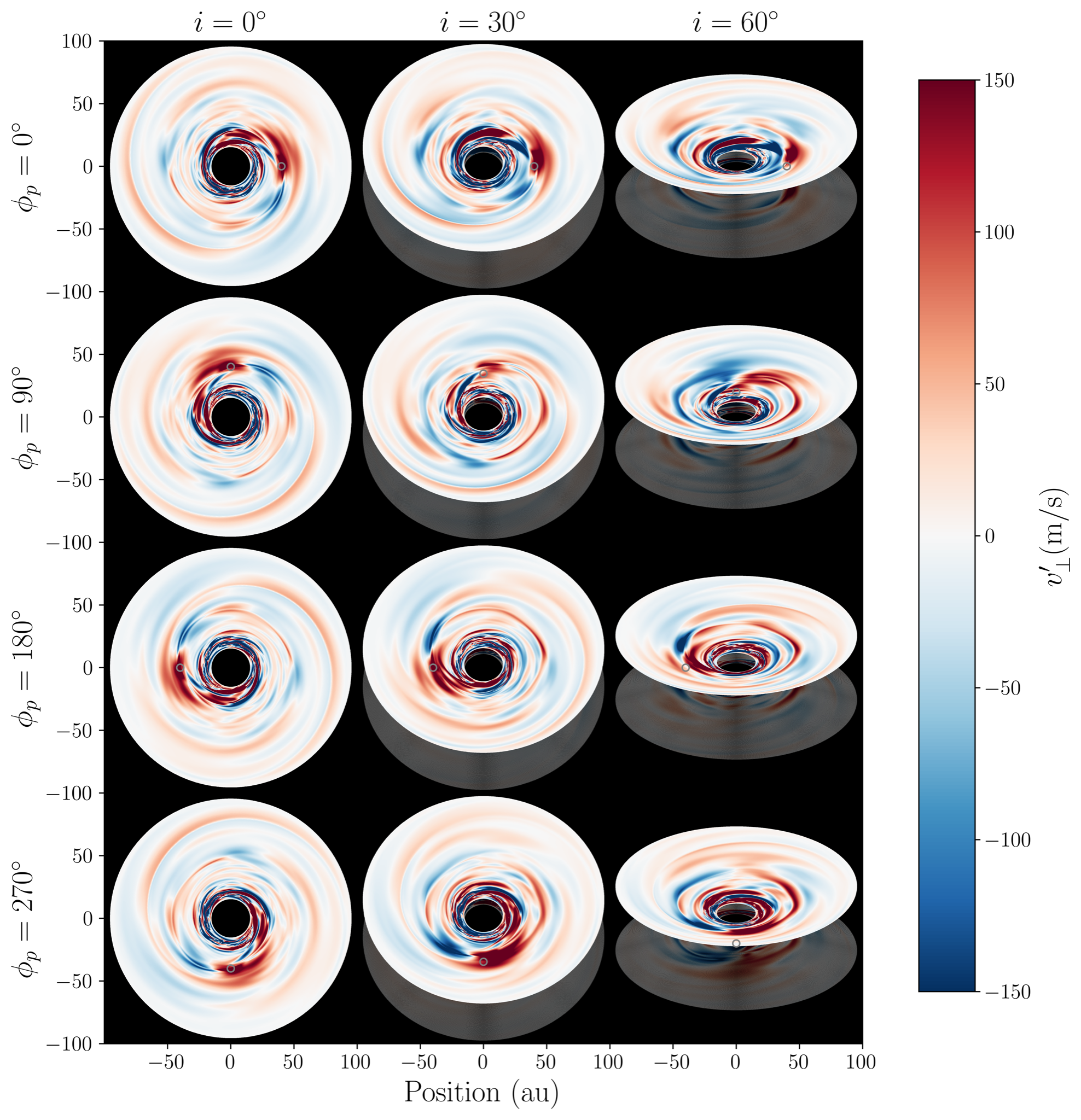}
    \caption{Sky-projected velocity perturbation $v'_\perp$ for a Jupiter-mass planet, with no accretion luminosity. Relative to the fiducial case, circumplanetary and in-gap flow patterns are significantly stronger.}
    \label{fig:jupiter_spv}
\end{figure*}

\begin{figure*}
    \centering
    \includegraphics[width=0.9\textwidth]{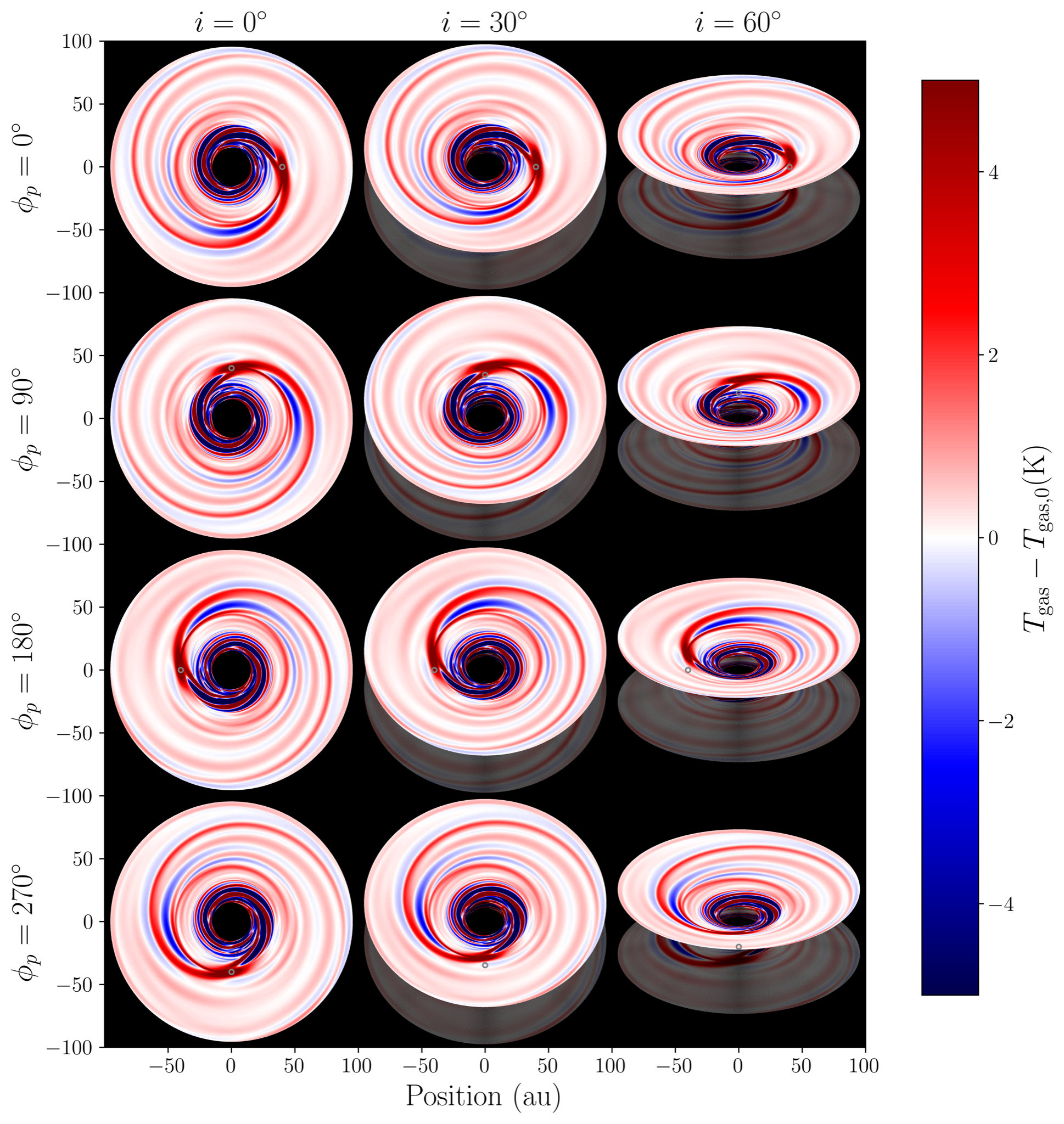}
    \caption{Perturbation in gas temperature $T_g - T_{g, 0}$ at $\theta = 0.3$ above the midplane, for our fiducial setup with a Saturn-mass, non-accreting planet. The double-armed structure of the Lindblad spiral in the upper atmosphere is clearly visible here.}
    \label{fig:3t_temp}
\end{figure*}

\begin{figure*}
    \centering
    \includegraphics[width=0.9\textwidth]{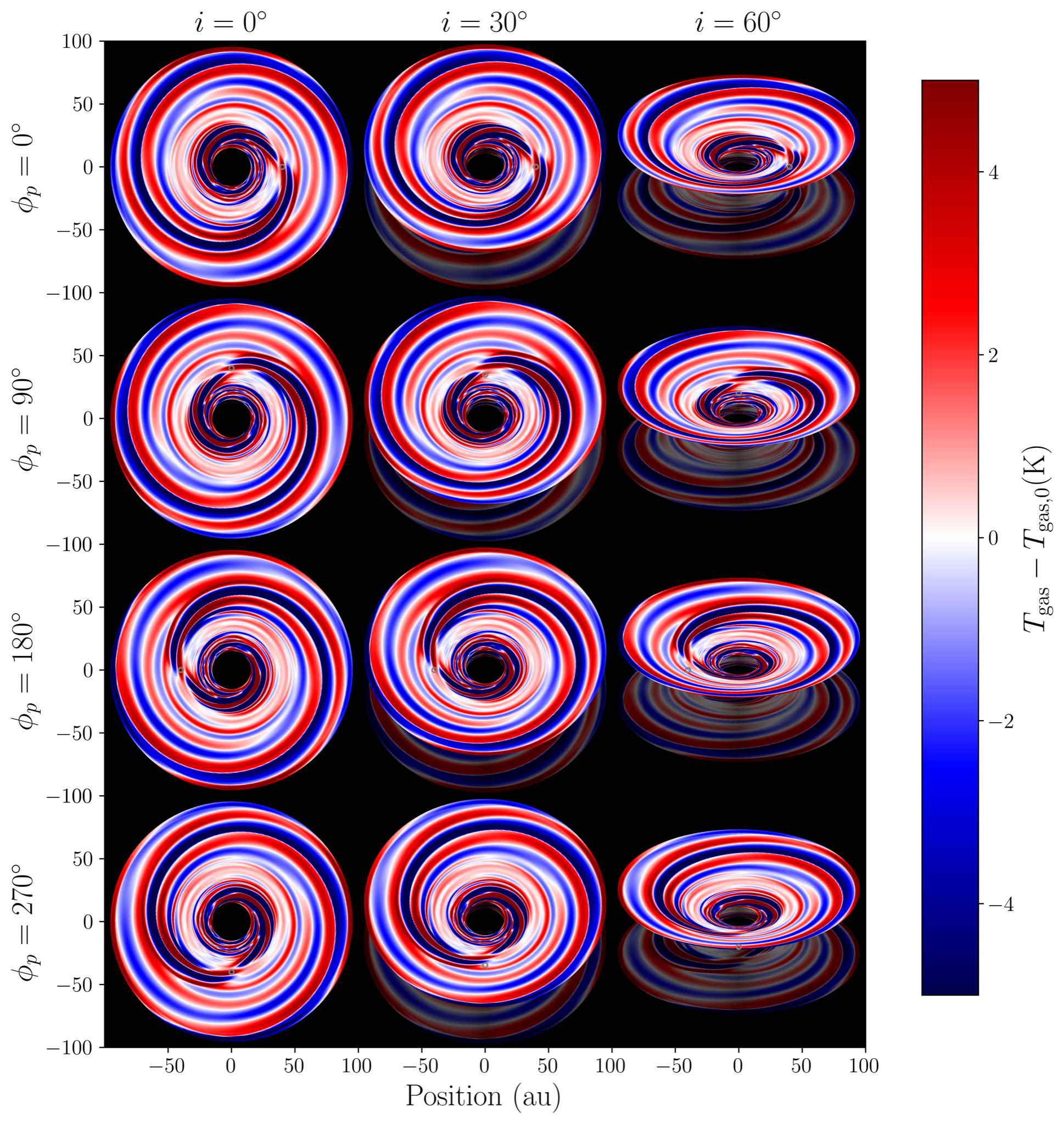}
    \caption{Gas temperature perturbation for an accreting, Saturn-mass planet. As with the kinematic spiral, the thermal spiral extends over a larger radial range and is more prominent than in the non-accreting case. }
    \label{fig:acclum_temp}
\end{figure*}

\begin{figure*}
    \centering
    \includegraphics[width=0.9\textwidth]{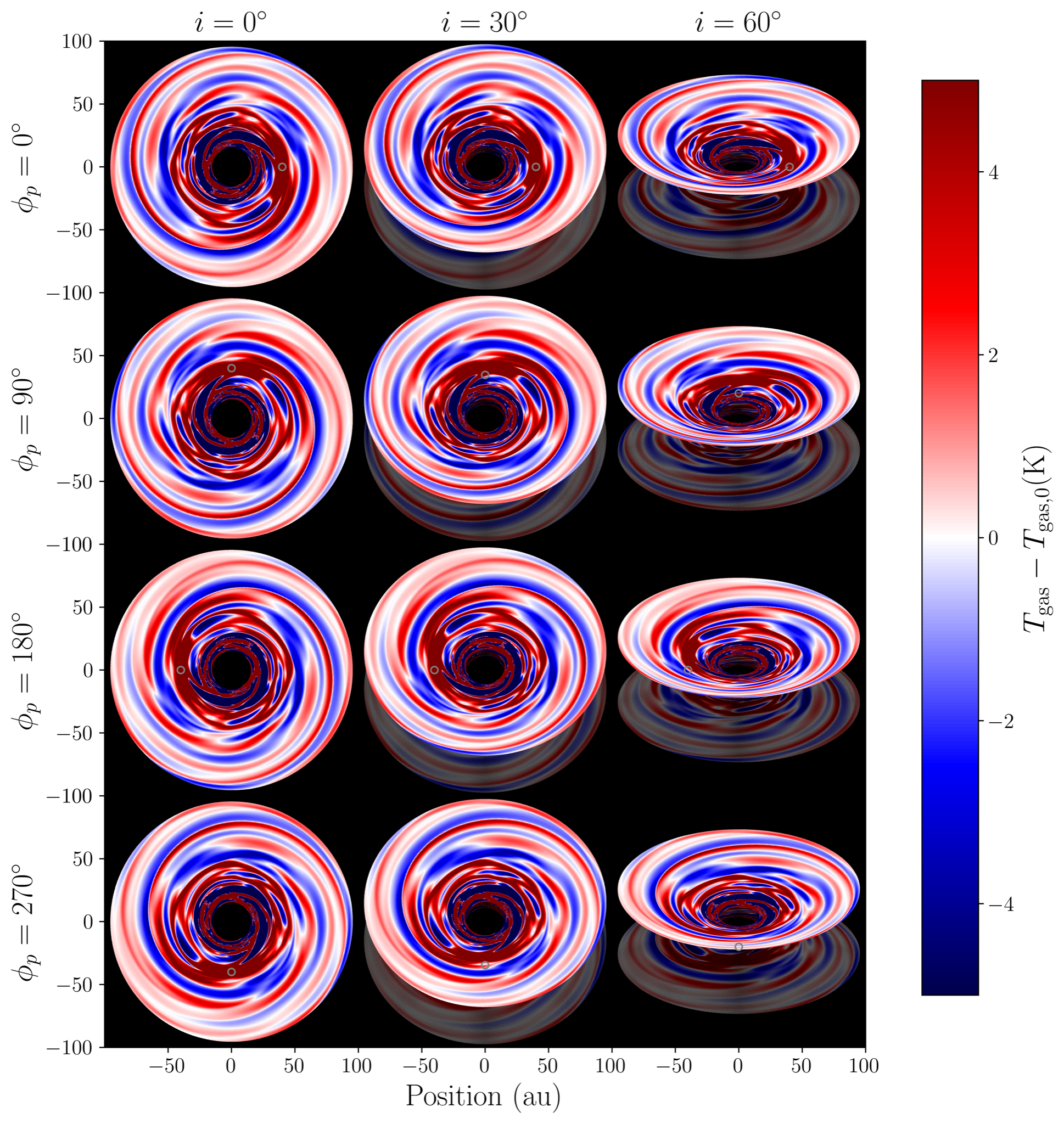}
    \caption{Gas temperature perturbation for a Jupiter-mass planet. As with kinematics, the circumplanetary region and gap make a much larger contribution to the thermal signature than in the fiducial Saturn-mass, non-accreting case.}
    \label{fig:jupiter_temp}
\end{figure*}

Among the cases we test, we find that (non-accreting) Saturn-mass planets, accreting Saturn-mass planets, and Jupiter-mass (non-accreting) planets are best at driving thermal and kinematic signatures that are amenable to observation. In each of these cases, we plot the planet-induced total velocity perturbation---projected along the line of sight ($v'_\perp$)---in Figures \ref{fig:3t_spv}, \ref{fig:acclum_spv}, and \ref{fig:jupiter_spv}, respectively. In Figures \ref{fig:3t_temp}, \ref{fig:acclum_temp}, and \ref{fig:jupiter_temp}, we likewise plot the perturbations in gas temperature $T_g$. In these figures, we take a cut at $\theta = 0.3$ above the midplane; this corresponds to $\sim 4.6$ scale heights at $r_p$, and roughly aligns with the expected $^{12}$CO J=$(2-1)$ emission layer \citep{BarrazaAlfaro2023}. (The $\theta = 0.2$ surface shown prominently in previous figures, at $\sim 3.1$ scale heights above the midplane, is closer to the $^12$CO J=$(3-2)$ layer.) We test disk inclinations $i_d = \{\ang{0}, \ang{30}, \ang{60}\}$ and planetary position angles $\phi_{p} = \{\ang{0}, \ang{90}, \ang{180}, \ang{270}\}$.

For an $i_d = \ang{0}$, $v'_\perp$ is dominated by $v'_{\theta}$. In the non-accreting, Saturn-mass fiducial case, the observed velocity spiral is driven by the local pressure-gradient term, as identified in Section \ref{sec:flow_analysis}; the redshifted spot above the planet location reflects the polar planetary flow shown in Figure \ref{fig:acclum_dens_cyl}. For the Saturn-mass planet with accretion luminosity, the outer Lindblad spiral is noticeably stronger and extends over a greater distance, while the planetary flow---no longer oriented in a polar direction---aligns with it, forming a double-banded velocity profile as discussed in Section \ref{sec:acclum}. In the Jupiter-mass case, the polar accretion flow is more prominent than in the fiducial case, with the spiral becoming subdominant. In observations, these features could help distinguish between prominent spirals created by high-mass planets and those created by highly luminous ones.

As disk inclination increases from zero, $v'_\perp$ increasingly reflects $v'_r$ (and to a lesser extent $v'_\phi$). Because the typical magnitude of $v'_r$ in spirals is typically much larger than $v'_{\theta}$, this strengthens the observed spiral signature, both in the inner and outer disks. Inclination also interacts with the elevation angle $\theta = 0.3$ of the plotted disk surface to distort the sky-projected areas of surface elements in an azimuthally dependent way, with disk patches on the near side of the star appearing smaller than those on the far side. Depending on the planetary position angle $\phi_p$, these properties can emphasize or deemphasize the circumplanetary region, inner, or outer spirals.

In each case, temperature spirals generally follow the same structure as the kinematic spirals; however, the magnitude of the observed temperature perturbation at a given point in the disk is a scalar quantity, and unlike $v'_\perp$ does not change with projection. Furthermore, as discussed in Sections \ref{sec:thermodynamic_test} and \ref{sec:planet_mass}, the nonzero, finite cooling time introduces a slight offset between the temperature and velocity perturbations. Conversely, the offset between kinematic and thermal spirals---as well as the prominence of buoyancy spirals---in different tracers probing various disk layers \citep[e.g., as studied in the MAPS program; ][]{Calahan2021MAPS} could help put limits on cooling time at various vertical positions in the disk. Planet-induced radial midplane shadows could also be used to measure cooling rates---by measuring their deviation from a straight line going through planet and star---provided that the planet's position angle is well-constrained. However, fully exploring these mechanisms would require a large hydrodynamical parameter study over disk masses and dust size distributions.

Our simulations run for a relatively short period of time, emphasizing the development of spiral density waves. This means that the effects of gap opening---in particular, changes to disk illumination and equilibrium temperature \citep{JangCondell2012} and changes to $v'_{\phi}$ due to the pressure gradient at gap edges \citep[][Section 2.4]{Armitage2020book}---have not had the time to fully develop. Especially in the Jupiter-mass case, these effects can be substantial. Simulations to gap-opening timescales are currently in progress, and we intend to present them---along with Monte Carlo radiative transfer (MCRT) post-processing and synthetic observations---in a forthcoming work.

\section{Conclusion}
We have run 3D, three-temperature radiation hydrodynamical simulations with the aim of better understanding the kinematic signatures that would be generated by forming protoplanets. Our fiducial setup, with a non-accreting Saturn-mass planet located at $r_p = 40 {\rm au}$ in a 15 $M_J$ disk, draws inspiration from the TW Hya system, where spiral arms---potentially excited by a forming planet \citep[e.g.,][]{Muley2021, Bae21}---have already been observed in temperature and velocity \citep[e.g.,][]{Teague2019}. First, we study the physical properties of the simulated planet-driven spirals, and compare our three-temperature approach to several thermodynamic prescriptions commonly used in the literature. Thereafter, we investigate the effects that changing planetary mass and accretion luminosity have on the strength and morphology of planet-driven disk features.

For our fiducial disk, we find that the results of our three-temperature simulations agree well with those from physically-motivated $\beta$-cooling. This naturally follows from the fact that for this setup the background equilibrium temperature, $T_{\rm eq}$, does not change much from the initial condition. As expected from previous works \citep[e.g.,][]{Juhasz2018, Muley2021, Bae21}, Lindblad spirals become more open with higher altitude in the stratified temperature background; buoyancy spirals are weak in strength, due to the relatively short cooling times we use. In the upper disk---most amenable to observation in $^{12}$CO---thermal and kinematic signatures are driven largely by local source terms, rather than transport across vertical gradients within the disk.

Our parameter survey shows that thermal and kinematic Lindblad spirals become stronger at high planet mass. Especially in the super-thermal mass regime, deeper planetary potential wells and larger Hill radii also enhance the signatures of circumplanetary flows. Planetary accretion luminosity adds pressure support to the circumplanetary region, and reorients the classic polar accretion flows \citep[e.g.,][]{Fung2015b,Fung2016,Fung2019} slightly outward. The separate effects of increasing planet mass versus accretion luminosity, relative to our fiducial case, are clearly visible in our sky-projected, inclined and rotated plots of the upper disk layers, and thus, potentially, in ALMA line observations.

Future areas of research could include testing different disk masses and dust-to-gas ratios, increasing resolution to allow for the development of hydrodynamical instabilities which may impact the visibility of spiral signatures \citep[e.g.,][]{BarrazaAlfaro2023}, or running simulations to gap-opening timescales \citep[e.g.,][]{FSC14, Fung2016} and post-processing the results to compare to observations; the last work is currently in progress.

More sophisticated numerical approaches would expand the scope of these comparisons. To more closely reproduce spiral pitch angles and morphologies in real disks \citep{Miranda2019a}, particularly for the 20-200 K temperature range spanned by our simulations, it would significantly help to relax the ideal-gas assumption, and compute $\gamma$ as a function of temperature, as well as hydrogen (para-, ortho-, atomic), helium, and metal fractions \citep{Boley2007,Bitsch2013,Boley2013}. Incorporating multiple dust species (including momentum exchange and turbulent diffusion) would enable modeling of widened, thickened rings of millimeter grains at planetary gap edges \citep{Bi2021,Bi2023}, visible in dust continuum, whereas a short-characteristics approach to radiation transport \citep[e.g.,][]{Davis2012}, which would allow for beam-crossing between accretion luminosity and reprocessed stellar radiation, could more accurately simulate strengths for radial shadows from luminous planets. Coupling such methods with our three-temperature scheme, however, is a computationally expensive and technically ambitious undertaking that we defer to future work.

\begin{acknowledgements}
We thank Myriam Benisty and Kees Dullemond for the suggestion to include sky-projected maps of velocity and temperature, as well as Marcelo Barraza Alfaro and Richard Teague for detailed comments on the manuscript. We also acknowledge useful discussions with Xue-Ning Bai, Jiaqing Bi, Remo Burn, Can Cui, Mario Flock, Dominik Ostertag, Jiahan Shi, Jess Speedie, and Alexandros Ziampras. We thank the anonymous referee for a helpful report, and particularly for encouraging us to comment on the equation of state. Numerical simulations were run on the Cobra and Raven clusters of the Max-Planck-Gesellschaft and the Vera Cluster of the Max-Planck-Institut f\"ur Astronomie, both hosted by the Max Planck Computing and Data Facility (MPCDF) in Garching bei München. The research of J.D.M.F. and H.K. is supported by the German Science Foundation (DFG) under the priority program SPP 1992: ``Exoplanet Diversity" under contract KL 1469/16-1/2. 

\end{acknowledgements}




\begin{appendix}
\section{Advective terms in spherical coordinates}\label{sec:adv_terms}
For a scalar quantity such as density or temperature, the rate of change at a specific location due to advection in any given direction is given by the (negative) gradient of the scalar, projected along the velocity in that direction. The advection of \textit{perturbations} is given by the advective term in the current state, minus advection in the initial condition. 

Making use of the fact that our initial condition is axisymmetric in all variables, and that $v_r(t = 0) = v_\theta(t = 0) \equiv 0$, we can write the evolution of the density perturbation $\rho'$ as:

\begin{subequations}
    \begin{equation}
    \left.\frac{\partial \rho'}{\partial t}\right|_{\rm adv r} = -v'_r \partial_{r}\rho
\end{equation}
\begin{equation}
    \left.\frac{\partial \rho'}{\partial t}\right|_{\rm adv \theta} = -\frac{v'_{\theta}}{r}\partial_{\theta}\rho
\end{equation}
\begin{equation}
    \left.\frac{\partial \rho'}{\partial t}\right|_{\rm adv \phi} = -\frac{v_{\phi}}{r \sin \theta }\partial_{\phi}\rho
\end{equation}
\end{subequations}

and that of the gas temperature perturbation $T_g'$ as
\begin{subequations}
\begin{equation}
    \left.\frac{\partial T_g'}{\partial t}\right|_{\rm adv r} = -v'_r \partial_{r}T_g
\end{equation}
    \begin{equation}
    \left.\frac{\partial T_g'}{\partial t}\right|_{\rm adv \theta} = -\frac{v'_{\theta}}{r}\partial_{\theta}T_g
\end{equation}
    \begin{equation}
    \left.\frac{\partial T_g'}{\partial t}\right|_{\rm adv \theta} = -\frac{v'_{\phi}}{r \sin \theta}\partial_{\phi}T_g
\end{equation}
\end{subequations}

For vectors, such as velocity, the advection term includes not only partial derivatives of each component in each coordinate, but also geometric connection terms arising from the change in the basis vectors themselves as a function of coordinate. This yields the following advective terms in each direction:
\begin{subequations}\label{eq:velocity_advection}
    \begin{equation}
    \left.\frac{\partial \vec{v}'}{\partial t}\right|_{\rm adv r} = -v'_r \left[\partial_r v'_r \hat{\vec{r}} + \partial_r v'_{\theta} \hat{\vec{\theta}} + \partial_r v_{\phi} \hat{\vec{\phi}} \right]
\end{equation}
    \begin{equation}
    \left.\frac{\partial \vec{v}'}{\partial t}\right|_{\rm adv \theta} = -\frac{v'_{\theta}}{r}\left[\left(\partial_{\theta} v'_r - v'_{\theta}\right) \hat{\vec{r}} + \left(\partial_{\theta} v'_{\theta} + v'_{r}\right) \hat{\vec{\theta}} + \left(\partial_{\theta} v'_{\phi} + \partial_{\theta} v_{0,\phi} \right) \hat{\vec{\phi}} \right]
\end{equation}
\begin{equation}
\begin{split}
    \left.\frac{\partial \vec{v}'}{\partial t}\right|_{\rm adv \phi} = &-\frac{v_{\phi}}{r \sin \theta}\left[\partial_{\phi} v'_r \hat{\vec{r}} + \partial_{\phi} v'_{\theta} \hat{\vec{\theta}} + \left(\partial_{\phi} v_{\phi} + v_r \sin \theta + v_{\theta} \cos \theta\right)\hat{\vec{\phi}}\right]\\
    & + \frac{v_\phi^2 - v_{\phi, 0}^2}{r \sin \theta} \left(\sin \theta \hat{\vec{r}} +\cos \theta \hat{\vec{\theta}}\right)
\end{split}
\end{equation}
\end{subequations}

We emphasize that despite considering only advection in one given coordinate direction, each of the expressions in Equation \ref{eq:velocity_advection} has three components, arising from the advection of velocity components orthogonal to the advection direction, as well as the aforementioned geometric connection terms. For ease of interpretation, in Figure \ref{fig:source_transport_outer}, we define the in-plane advection as the sum $\left(\partial \vec{v}'/\partial t\right)_{\rm adv, in-plane} = \left(\partial \vec{v}'/\partial t\right)_{\rm adv r} + \left(\partial \vec{v}'/\partial t\right)_{\rm adv\phi}$.
\vfill
\end{appendix}

\end{document}